\documentclass[review]{elsarticle}

\usepackage{lineno,hyperref}
\usepackage{amssymb}
\usepackage{color}
\usepackage{amsmath}
\usepackage{tikz}
\usepackage{amsfonts}
\usepackage{mathrsfs}
\usepackage{amsthm}

\usepackage{bbding}
\usepackage{mathrsfs}
\usepackage{amsmath, amsfonts, amssymb, mathrsfs, txfonts}
\usepackage{graphicx, subfigure}
\usepackage{wasysym}
\usepackage{color}
\usepackage{bm}
\usepackage{enumerate}

\usepackage{pifont}
\usepackage{txfonts}
\usepackage{amsmath}
\usepackage{graphicx}
\usepackage{amsfonts}
\usepackage{amssymb}
\usepackage{mathrsfs,psfrag,eepic,epsfig}
\usepackage{epstopdf}

 \theoremstyle{definition}
 
 \theoremstyle{remark}

\biboptions{numbers,sort&compress}

\modulolinenumbers[5]

\journal{Journal of \LaTeX\ Templates}

\makeatletter \@addtoreset{equation}{section}

\begin{document}

\begin{frontmatter}
\title{Riemann-Hilbert problem for the modified Landau-Lifshitz equation with nonzero boundary conditions}
\tnotetext[mytitlenote]{Project supported by the Fundamental Research Fund for the Central Universities under the grant No. 2019ZDPY07.\\
\hspace*{3ex}$^{*}$Corresponding author.\\
\hspace*{3ex}\emph{E-mail addresses}: jinjieyang@cumt.edu.cn (J.J. Yang), sftian@cumt.edu.cn,
shoufu2006@126.com (S. F. Tian)}

\author{Jin-Jie Yang and Shou-Fu Tian$^{*}$}
\address{
School of Mathematics and Institute of Mathematical Physics, China University of Mining and Technology,\\ Xuzhou 221116, People's Republic of China\\
}

\begin{abstract} We study systematically a matrix Riemann-Hilbert problem for the modified Landau-Lifshitz (mLL) equation with nonzero boundary conditions at infinity. Unlike the zero boundary conditions case, there occur double-valued functions during the process of the direct scattering.  In order to establish the Riemann-Hilbert (RH) problem, it is necessary to make appropriate modification, that is, to introduce an affine transformation that can convert the Riemann surface into a complex plane. In the direct scattering problem, the analyticity, symmetries, asymptotic behaviors of Jost functions and scattering matrix are presented in detail. Furthermore, the discrete spectrum, residual conditions, trace foumulae and theta conditions are established with simple and double poles. The inverse problems are solved via a matrix RH problem formulated by Jost function and scattering coefficients.  Finally, the dynamic behavior of some typical soliton solutions of the mLL equation with reflection-less potentials are given to further study the structure of the soliton waves.   In addition, some remarkable characteristics of these soliton solutions are
analyzed graphically. According to analytic solutions, the influences of each parameters on dynamics of the soliton
waves and breather waves  are discussed, and the method of how to control such nonlinear phenomena are
suggested.
\end{abstract}


\end{frontmatter}

\linenumbers

\section{Introduction}
Nonlinear Schr\"{o}dinger equations are significant  mathematical and physical model. It has been applied in many fields, such as magnetic field, optics etc. However, it is not enough to describe the complexity of the phenomena only by the nonlinear Schr\"{o}dinger equation, which leads to some specific terms such as Sasa-Satsuma equation, Hirota equation etc need to be added.  In electromagnetics, the mLL equation \cite{Bazaliy,Slonczewski} can be used to describe the dynamic behavior of local magnetization
\begin{align}\label{Q2}
\frac{\partial\mathbf{M}}{\partial t^{'}}=-\gamma\mathbf{M}\times\mathbf{M}_{eff}+
\frac{\rho}{M_{s}}\times\frac{\partial\mathbf{M}}{\partial t^{'}}+\mathbf{\tau}_{b},
\end{align}
here $\gamma$ is the gyromagnetic ratio, $\rho$ stands the Gilbert damping parameter,  the localized magnetization $\mathbf{M}\equiv \mathbf{M}(x,t)$ and $\mathbf{M}_{eff}$ implies the effective magnetic field such as  the external field, the anisotropy field, the demagnetization field,
and the exchange field which is  equivalent to
\begin{align}\label{L1}
\mathbf{M}_{eff}=\frac{2A}{M_{s}^{2}}\frac{\partial^{2}\mathbf{M}}{\partial x^{2}}+
[(\frac{H_{k}}{M_{s}}-4\pi)M_{z}+H_{ext}]\mathbf{e}_{z},
\end{align}
where $A$ is the exchange constant, $H_{k}$ denotes the anisotropy field, $H_{ext}$ expresses the external field, $H_{z}$ is the demagnetization field, as well as  $\mathbf{e}_{z}$ shows the unit vector along the $z$ direction. Now letting the vector $\mathbf{m}=\mathbf{M}/M_{s}$, one has the another expression of \eqref{Q2}
as  follow
\begin{align}\label{L2}
\frac{\partial\mathbf{m}}{\partial t}=
-\mathbf{m}\times\frac{\partial^{2}\mathbf{m}}{\partial x^{2}}+
\delta\mathbf{m}\times\frac{\partial\mathbf{m}}{\partial t}+
\frac{b_{J}t_{0}}{l_{0}}\frac{\partial\mathbf{m}}{\partial x}-
(m_{z}+\frac{H_{ext}}{H_{k}-4\pi M_{s}})\mathbf{m}\times\mathbf{e}_{z},
\end{align}
with $t=t^{'}/t_{0}$, $x=x^{'}/l_{0}$, $t_{0}=1/[\gamma(H_{k}-4\pi M_{s})]$ expresses the characteristic time, $l_{0}=\sqrt{2A/[M_{s}(H_{k}-4\pi M_{s})]}$ denotes  the characteristic length. Obviously, when $\mathbf{m}\equiv(m_{x},m_{y},m_{z})=(0,0,1)$,  the
ground state of the system, and two kinds of nonlinear excited state can be obtained, which means the spin-wave solution and magnetic soliton. Next, we consider that the magnetic field is large enough, and the deviation of the magnetization of the two excited states from the ground state is small. It is reasonable to take  a reasonable transformation
\begin{align}
\hat{q}=m_{x}+im_{y},\quad m_{z}=\sqrt{1-|\hat{q}|^{2}}.
\end{align}
Substituting the  equation into \eqref{L2} yields
\begin{align}\label{Xin}
i\frac{\partial\hat{q}}{\partial t}-m_{z}\frac{\partial^{2}\hat{q}}{\partial x^{2}}
+\hat{q}\frac{\partial^{2}m_{z}}{\partial x^{2}}+\delta\left(
m_{z}\frac{\partial\hat{q}}{\partial t}-\hat{q}\frac{\partial m_{z}}{\partial t}
\right)-i\frac{b_{J}t_{0}}{l_{0}}\frac{\partial\hat{q}}{\partial x}
+\left(m_{z}+\frac{H_{ext}}{H_{k}-4\pi M_{s}}\right)=0.
\end{align}

In this work, different kinds of soliton solutions   under non-zero boundary conditions are expected to be obtained via the RH problem in a uniaxial ferromagnetic nanowire with spin torque. However \eqref{Xin} is not integrable. To achieve this goal, we consider the case without damping and the long wave length approximation \cite{Kosevich}.  Keeping only the nonlinear terms of
the order of magnitude of $|\hat{q}|^{2}\hat{q}$, then the equation \eqref{Xin} can be written as the following integrable equation
\begin{align}\label{L3}
i\hat{q}_{t}-\hat{q}_{xx}-\frac{1}{2}|\hat{q}|^{2}\hat{q}+
(1+\frac{H_{ext}}{H_{k}-4\pi M_{s}})\hat{q}
-i\frac{b_{J}t_{0}}{l_{0}}\hat{q}_{x}=0,
\end{align}
Much work has been done on this equation \eqref{L3} at home and abroad. The soliton solutions with the  Hirota method \cite{Hirota-1}-\cite{Hirota-3}; Conservation laws, modulation instability and rogue waves \cite{Su-2017};   To discuss the properties of the soliton solution on the spin-wave background,  Li and his team use a straightforward Darboux transformation \cite{DT}  to construct the general expression for the soliton solution \cite{Li-2007}; Rogue waves for \eqref{L3} has been studied \cite{Zhao}, which presents that the accumulation of energy  plays the vital role for the generation of the magnetic rogue waves etc. Recently, the inverse scattering (IS) theory of nonlinear equations under non-zero boundary conditions  \cite{NZBC-1}-\cite{NZBC-12} based on the construction of RH \cite{RHP-1}-\cite{RHP-7}problem has attracted considerable attention.    In fact, the initial value problem for integrable models can generally be solved by the IS theory, a special situation of Fourier analysis. Since the IS method reconstructing the solution form  was proposed to solve Cauchy problem of the integrable KdV model by Gardner, Greene, Kruskal, Miura \cite{GGKM}, which is widely applied to other equations such as Camassa-Holm equation \cite{RH-1}, Kundu-Eckhaus equation \cite{RH-2}, Gerdjikov-Ivanov type of derivative nonlinear Schr\"{o}dinger equation \cite{RH-3}, coupled mKdV system \cite{RH-4}, a generalized Sasa-Satsuma equation \cite{RH-5}, Maxwell-Bloch equations \cite{RH-6} etc.

 But as far as we know, the  IS theory and soliton solution for  equation \eqref{L3} under non-zero boundary conditions have not been reported. There are many problems that need to be solved, such as how to  introduce affine transformation in order to overcome the multiplicity of eigenfunctions, how to transform appropriately the function boundary condition $qe^{(\delta_{1}-2iq_{0}^{2})t}$ to  constant one $q$, how to   introduce an invertible matrix in order to study the spectral problem of asymptotic Lax pair transformed into diagonal matrix problem, how to study the analyticity of Jost function and scattering matrix on these basis. The main purpose of our work is to study these problem in order to construct   the generalized RH problem of equation \eqref{L3} by using the Jost functions and scattering data. Then the residue conditions of discrete spectral points are  analyzed, and the expression of the solution of   equation \eqref{L3}  is also obtained under simple poles and double poles conditions. In addition, the trace formulas and theta conditions of simple poles and double poles are different from.

The outline of the work is arranged as:
In section 2, the  affine transformation is introduced to transform the boundary conditions into constants, and then the asymptotic Lax pairs are obtained. Furthermore, the analytic and symmetric properties of Jost functions and scattering matrix are obtained by spectral analysis. The residue conditions are given in order to study the inverse transformation process. In section 3, we study the RH problem at a single-pole and reconstruct its potential function. In section 4, we discuss the propagation behavior of different kinds for  solutions by choosing appropriate parameters with a brief analysis. The construction of solutions  with doubles poles, trace formula and theta condition are also studied in section 5. Finally, some conclusions and discussions are presented in the last section.

\section{Direct scattering problem with NZBCs}
In this section, the analyticity and asymptotic of the Jost function, asymptotic of the scattering matrix, symmetries, discrete spectrum, and residue conditions will be presented during the direct scattering. In addition, different from the zero boundary value problem, multi-valued function will appear in the process of calculation, we need to introduce appropriate transformation to convert multi-valued function into single-valued function to facilitate the research.
\subsection{Lax pair of mLL}
Considering the Lax pair of the mLL equation \eqref{L2} and
letting $q=\hat{q}/2$, with the NZBCs as $x\rightarrow\pm\infty$
 \begin{align}
 \lim \mathop{q(x,y)}_{x\rightarrow\pm\infty}= qe^{(\delta_{1}-2iq_{0}^{2})t},
 \end{align}
 one has the equivalent Lax pair
\begin{align}\label{QQ1}
&\left\{ \begin{aligned}
&\phi_{x}=X\phi,\quad X=ik\sigma_{3}+Q,\\
&\phi_{t}=T\phi, \quad T=[2ik^{2}+i\delta_{2}k+\frac{1}{2}
(\delta_{1}-2i|q|^{2})+iQ_{x}]\sigma_{3}+2kQ+\delta_{2}Q,
     \end{aligned}  \right.
\end{align}
with
\begin{align*}
Q=\left(\begin{array}{cc}
    0  & q \\
    -q^{*} & 0  \\
  \end{array}\right),\qquad
  \sigma_{3}=\left(\begin{array}{cc}
    1  & 0 \\
    0 & -1  \\
  \end{array}\right),
  \end{align*}
where $k$ is an spectrum parameter, the superscript $*$ represents the complex conjugate, $\delta_{1}=1+\frac{H_{ext}}{H_{k}-4\pi M_{s}}$, $\delta_{2}=i\frac{b_{J}t_{0}}{l_{0}}$ and the function $\phi$ is a $2\times2$ matrix.

\noindent \textbf {Theorem 1.}
\emph{Let's introduce the transformation}
\begin{align}
\begin{split}
&q= qe^{(\delta_{1}-2iq_{0}^{2})t},\\
&\phi= e^{-\frac{i}{2}(2q_{0}^{2}+i\delta_{1})t\sigma_{3}}\psi,
\end{split}
\end{align}
\emph{then  its boundary values changes to $q(x,t)\rightarrow q_{\pm}$ as $x\rightarrow\pm\infty$.}

Finally one can obtain the following asymptotic scattering problem under Theorem $1$ of the Lax pair for the mLL equation at infinity
\begin{align}\label{Q3}
&\left\{ \begin{aligned}
&\psi_{x}=X_{\pm}\psi,\quad X_{\pm}=\lim_{x\rightarrow\pm\infty}X=ik\sigma_{3}+Q_{\pm},\\
&\psi_{t}=T_{\pm}\psi, \quad T_{\pm}=\lim_{x\rightarrow\pm\infty}T=(2k+\delta_{2})X_{\pm},
     \end{aligned}  \right.
\end{align}
here $Q_{\pm}=\mathop{\lim}\limits_{x\rightarrow\pm\infty}Q=\left(\begin{array}{cc}
    0  & q_{\pm} \\
    -q^{*}_{\pm} & 0  \\
  \end{array}\right).$
\subsection{Riemann surface and uniformization coordinate}
Obviously the eigenvalue of asymptotic matrix $X_{\pm}$ are multi-valued function $\pm i\sqrt{k^{2}+q_{0}^{2}}$, and in this case, unlike the zero boundary value, to deal with this situation we need to introduce a two-sheeted Riemann surface defined by
\begin{align}\label{Q4}
\lambda^{2}=k^{2}+q_{0}^{2},
\end{align}
where the two-sheeted Riemann surface completed by gluing together two copies of extended complex $k$-plane $S_{1}$ and $S_{2}$ along the cut $iq_{0}[-1,1]$ between the branch points $k=\pm iq_{0}$ obtained by the value of $\sqrt{k^{2}+q_{0}^{2}}=0$.
Introducing the local polar coordinates
\begin{align}\label{Q5}
k+iq_{0}=r_{1}e^{i\theta_{1}},\quad k-iq_{0}=r_{2}e^{i\theta_{2}},\quad -\frac{\pi}{2}<\theta_{1},\theta_{2}<\frac{3\pi}{2},
\end{align}
we get a single-valued analytical function on the Riemann surface
\begin{align}\label{Q6}
\lambda(k)=&\left\{\begin{aligned}
&(r_{1}r_{2})^\frac{1}{2}e^\frac{{\theta_{1}+\theta_{2}}}{2}, \quad &on\quad S_{1},\\
-&(r_{1}r_{2})^\frac{1}{2}e^\frac{{\theta_{1}+\theta_{2}}}{2}, \quad &on\quad S_{2}.
\end{aligned} \right.
\end{align}
Define the uniformization variable $z$ by the conformal mapping \cite{LD}
\begin{align}\label{Q7}
z=k+\lambda,
\end{align}
and form \eqref{Q4}, one can get two single-value function
\begin{align}\label{Q8}
k(z)=\frac{1}{2}(z-\frac{q_{0}^{2}}{z}),\quad \lambda(z)=\frac{1}{2}(z+\frac{q_{0}^{2}}{z}).
\end{align}

\noindent \textbf {Theorem 2.}
\emph{According to conformal mapping \eqref{Q7}, some propositions can be observed as follows \\
 $\blacktriangleright$ $Im k>0$ of sheet $S_{1}$ and $Im k<0$ of sheet $S_{2}$ are mapped into $Im \lambda >0$;\\
 $\blacktriangleright$ $Im k<0$ of sheet $S_{1}$ and $Im k>0$ of sheet $S_{2}$ are mapped into $Im \lambda <0$;\\
 $\blacktriangleright$ The   branch  $[-iq_{0},iq_{0}]$ of $k$-plane is mapped into the branch  $[-q_{0},q_{0}]$ of $\lambda$-plane;\\
 $\blacktriangleright$ By the Joukowsky transformation map:\\
$\bullet$ $Im \lambda>0$ into domain}
\begin{align*}
D^{+}=\left\{z\in \mathbb{C}:\left(|z|^{2}-q_{0}^{2}\right)Im z>0\right\},
\end{align*}
\emph{which means the upper half of the $\lambda$-plane maps to the upper outer half of the circle of radius $q_{0}$ and the inner half of the circle of the lower half of the $z$-plane}.\\
$\bullet$ \emph{ $Im \lambda<0$ into domain}
\begin{align*}
D^{-}=\left\{z\in \mathbb{C}:\left(|z|^{2}-q_{0}^{2}\right)Im z<0\right\},
\end{align*}
\emph{which stands for the lower half of the $\lambda$-plane maps to the upper inner half of the circle of radius $q_{0}$ and the outer half of the circle of the lower half of the $z$-plane}.\\
$\blacktriangleright$ \emph{On the sheet $S_{1}$, $z\rightarrow\infty$ as $k\rightarrow\infty$; on the sheet $S_{2}$,  $z\rightarrow 0$ as $k\rightarrow\infty$.}

The results above can be summarized as the following picture
\\

\centerline{\begin{tikzpicture}[scale=0.5]
\filldraw[yellow, line width=0.5](1,0)--(3,0) arc (-180:0:2);
\path [fill=yellow] (1,0) -- (9,0) to
(9,4) -- (1,4);
\path [fill=yellow] (-8,0) -- (0,0) to
(0,4) -- (-8,4);
\filldraw[white, line width=0.5](3,0)--(7,0) arc (0:180:2);
\draw[fill] (5,0)node[below]{} circle [radius=0.035];
\draw[fill] (-4,-0)node[below]{} circle [radius=0.035];
\draw[-][thick](-8,0)--(-7,0);
\draw[-][thick](-7.0,0)--(-6.0,0);
\draw[-][thick](-6,0)--(-5,0);
\draw[-][thick](-5,0)--(-4,0)node[below right]{\footnotesize$0$};
\draw[-][thick](-4,0)--(-3,0);
\draw[-][thick](-3,0)--(-2,0);
\draw[-][thick](-2,0)--(-1,0);
\draw[->](-1,0)--(0,0)[thick]node[above]{$Rek$};
\draw[-][thick](-4,-4)--(-4,-3);
\draw[-][thick](-4,-3)--(-4,-2)node[below right]{\footnotesize$-iq_{0}$};
\draw[-][thick](-4,-2)--(-4,-1);
\draw[-][thick](-4,-1)--(-4,0);
\draw[-][thick](-4,0)--(-4,1);
\draw[-][thick](-4,1)--(-4,2)node[above right]{\footnotesize$iq_{0}$};
\draw[-][thick](-4,2)--(-4,3);
\draw[->][thick](-4,3)--(-4,4)[thick]node[right]{$Imk$};
\draw[->][thick](1,0)--(2,0);
\draw[-][thick](2,0)--(3,0);
\draw[-][thick](3,0)--(4,0);
\draw[<-][thick](4,0)--(5,0)node[below right]{\footnotesize$0$};
\draw[-][thick](5,0)--(6,0);
\draw[<-][thick](6,0)--(7,0);
\draw[->][thick](7,0)--(8,0);
\draw[-](8,0)--(9,0)[thick]node[above]{$Rez$};
\draw[-][thick](5,-4)--(5,-3);
\draw[-][thick](5,-3)--(5,-2)node[below right]{\footnotesize$-iq_{0}$};
\draw[-][thick](5,-2)--(5,-1);
\draw[-][thick](5,-1)--(5,0);
\draw[-][thick](5,0)--(5,1);
\draw[-][thick](5,1)--(5,2)node[above right]{\footnotesize$iq_{0}$};
\draw[-][thick](5,2)--(5,3);
\draw[-][thick](5,3)--(5,4)[thick]node[right]{$Imz$};
\draw[fill] (7,0) circle [radius=0.055][below right][thick]node{\footnotesize$0^{+}$};
\draw[fill] (3,0) circle [radius=0.055][below right][thick]node{\footnotesize$0^{-}$};
\draw[fill] (5,2) circle [radius=0.055];
\draw[fill] (5,-2) circle [radius=0.055];
\draw[fill] (-4,2) circle [radius=0.055];
\draw[fill] (-4,-2) circle [radius=0.055];
\draw[fill] (8,2.5) circle [radius=0.035][thick]node[right]{\footnotesize$z_{n}$};
\draw[fill][red] (8,-2.5) circle [radius=0.035][thick]node[right]{\footnotesize$z_{n}^{*}$};
\draw[fill][red] (3.7,1) circle [radius=0.035][thick]node[right]{\footnotesize$-\frac{q_{0}^{2}}{z_{n}}$};
\draw[fill] (3.7,-1) circle [radius=0.035][thick]node[right]{\footnotesize$-\frac{q_{0}^{2}}{z_{n}^{*}}$};
\draw[fill] (-2.5,1.5) circle [radius=0.035][thick]node[right]{\footnotesize$z_{n}$};
\draw[fill][red] (-2.5,-1.5) circle [radius=0.035][thick]node[right]{\footnotesize$z_{n}^{*}$};
\draw[-][line width=0.8] (7,0) arc(0:360:2);
\draw[->][line width=0.8] (7,0) arc(0:220:2);
\draw[->][line width=0.8] (7,0) arc(0:330:2);
\draw[->][line width=0.8] (7,0) arc(0:-330:2);
\draw[->][line width=0.8] (7,0) arc(0:-220:2);
\end{tikzpicture}}
\noindent {\small \textbf{Figure 1.} (Color online) Left Fig: the first sheet of the Riemann surface, presenting different discrete spectral points in the $k$-plane with $Imk>0$ (yellow) and $Imk<0$ (white); Right Fig: shows the discrete spectral points on the $z$-plane after introducing the transformation with $Imz>0$ (yellow) and $Imz<0$ (white), and gives  the orientation of the jump contours about the RH problem, where the red spectral points represent the zeros of $s_{11}(z)$ and the black spectral points represent the zeros of $s_{22}(z)$. }
\\

\subsection{Jost function and its analyticity}
Resort to the asymptotic Lax pair \eqref{Q3}, a invertible matrix $E_{\pm}$, asymptotic eigenvector matrix, can be obtained to  diagonalize the matrix of $X_{\pm}$ and $T_{\pm}$, namely
\begin{align}\label{Q12}
\begin{split}
  X_{\pm}(x,t;z)&=E_{\pm}(z)(i\lambda \sigma_{3})E_{\pm}^{-1}(z),\\
  T_{\pm}(x,t;z)&=E_{\pm}(z)[i\lambda(2k^{2}+\delta_{2})\sigma_{3}]E_{\pm}^{-1}(z).
  \end{split}
\end{align}

As usual, the continuous spectrum $\Sigma_{k}$ is composed of all $k$ values in both planes ($S_{1}$, $S_{2}$) and satisfies $\lambda_{k}\in\mathbb{R}$, namely, $\Sigma_{k}=\mathbb{R}\cup iq_{0}[-1,1]$. By mapping \eqref{Q7} the continuous spectrum becomes $\Sigma_{z}=\mathbb{R}\cup C_{0}$, where $C_{0}$ is a circle of radius $q_{0}$. For convenience, let's write $\Sigma_{z}$ as $\Sigma$. Now for all $z\in\Sigma$, the simultaneous solutions $\phi_{\pm}$ of the Lax pair \eqref{QQ1} can be formulated, which is called the Jost function, and caters
\begin{align}\label{Q13}
\phi_{\pm}(x,t;z)\thicksim\psi_{\pm}(x,t;z)=E_{\pm}(z)e^{i\theta(x,t;z) \sigma_{3}},
\quad x\rightarrow \pm\infty,
\end{align}
with $\theta(x,t;z)=\lambda(z)[x+(2k(z)+\delta_{2})t]$ and
\begin{align}\label{IV}
E_{\pm}(z)=\left(
  \begin{array}{cc}
     1 & \frac{iq_{\pm}}{k+\lambda} \\
     \frac{iq_{\pm}^{*}}{k+\lambda} & 1 \\
  \end{array}
\right)=\mathbb{I}+(i/z)\sigma_{3}Q_{\pm}.
\end{align}
Let
\begin{align}\label{TE}
u_{\pm}(x,t;z)=\phi_{\pm}(x,t;z)e^{-i\theta(x,t;z)\sigma_{3}},
\end{align}
which means
\begin{align}\label{Q14}
\lim_{x\rightarrow\pm\infty}u_{\pm}(x,t;z)=E_{\pm}(z).
\end{align}
Then the Lax pair of $u_{\pm}$ can be expressed as
\begin{equation}\label{Q15}
\begin{split}
&[E_{\pm}^{-1}(z)u_{\pm}(z)]_{x}+i\lambda [E_{\pm}^{-1}(z)u_{\pm}(z),\sigma_{3}]=E_{\pm}^{-1}(z)\Delta Q_{\pm}(z)u_{\pm}(z),\\
&[E_{\pm}^{-1}(z)u_{\pm}(z)]_{t}+i\lambda  (2k+\delta_{2})[E_{\pm}^{-1}(z)u_{\pm}(z),\sigma_{3}]=E_{\pm}^{-1}(z)\Delta T_{\pm}(z)u_{\pm}(z),
\end{split}
\end{equation}
here the subscript shows the partial derivatives with respect to $x$ and $t$, $\Delta Q_{\pm}(z)=Q-Q_{\pm}$ and  $\Delta T_{\pm}(z)=T-T_{\pm}$. Further the Lax pair \eqref{Q15} can be written in full derivative form.
So we take two particular paths, i.e., $(-\infty, x)$ and $(x, \infty)$, then the ODEs for $u_{\pm}$ can be determined by the following two eigenfunctions
\begin{align}\label{Q16}
\begin{matrix}
u_{-}(x,t;z)=E_{-}+\int_{-\infty}^{x}E_{-}e^{i\lambda(x-y)\sigma_{3}}E_{-}^{-1}\Delta Q_{-}(y,t)u_{-}(y,t;z)e^{-i\lambda(x-y)\sigma_{3}}\, dy,\\
u_{+}(x,t;z)=E_{+}-\int_{x}^{\infty}E_{+}e^{i\lambda(x-y)\sigma_{3}}E_{+}^{-1}\Delta Q_{+}(y,t)u_{+}(y,t;z)e^{-i\lambda(x-y)\sigma_{3}}\, dy.
\end{matrix}
\end{align}
Note that if $q(x)-q\in L^{1}(-\infty, a)$ for some $a\in \mathbb{R}$, then the Neumann series converges absolutely and uniformly with respect to $x\in(-\infty, a)$ and $z\in D_{\varepsilon}^{-}$ for all $\varepsilon >0$. And because a uniformly convergent series is an analytic function, it must converge to an analytic function \cite{M.J.,Henrici}, which shows that the corresponding column of the eigenfunction is analytic in this domain (for details, please refer to \cite{Biondini}). Then the
analyticity of the function $u_{\pm}$ can be obtained.

\noindent \textbf {Proposition 1.}
\emph{The columns $u_{+,1}(x,t;z)$ and $u_{-,2}(x,t;z)$ are analytically in $D_{+}$ of $z$-plane, and the columns $u_{-,1}(x,t;z)$ and $u_{+,2}(x,t;z)$ are analytically in $D_{-}$ of $z$-plane, here $u_{\pm,i}(x,t;z)$ ($i=1, 2$) denotes the $i$-th column of $u_{\pm}$}.

\noindent \textbf {Corollary 1.}
\emph{Assume that the unique solution $\phi_{\pm}(x,t;z)$ to \eqref{L1} can be expressed by \eqref{Q13}, then the analyticity of Jost function $\phi_{\pm}(x,t;z)$ is consistent with that of the function $u_{\pm}(x,t;z)$, i.e., $\phi_{+,1}(x,t;z)$ and $\phi_{-,2}(x,t;z)$ are analytically in $D_{+}$ of $z$-plane, one the other hand,  $\phi_{+,2}(x,t;z)$ and $\phi_{-,1}(x,t;z)$ are analytically in $D_{-}$ of $z$-plane as well as $\phi_{\pm,i}(x,t;z)$ ($i=1, 2$) represent the $i$-th column of $\phi_{\pm}$}.

\noindent \textbf {Theorem 3.} (\textbf{Liouville's formula})
\emph{Let's say that $M$ is a $n$-th order matrix and satisfies a homogeneous linear differential equation $Y'=M(x)Y$, here $Y$ is an $n$ vector. If a matrix $M$ is a solution of the differential equation,  one has $(\det Y)_{x}=trM\det Y$, further $\det Y(x)=\det Y(x_{0})e^{\int_{x_{0}}^{x}trM(\zeta)d\zeta}$. }

\noindent \textbf {Corollary 2.}
\emph{Suppose that the matrix $M$ where the trace is equal to zero satisfies a linear differential equation defined by Theorem 3, thus we can get $(\det Y)_{x}=0$.}
\subsection{Scattering matrix, scattering coefficients and reflection coefficients}
In this subsection, we discuss the dependence of Jost functions $\phi_{\pm}$ and obtain the reflection coefficients. Recall the expression for $X(z)$ and $T(z)$ of Lax pair defined by \eqref{L1}, one has $(\det\phi)_{x}=(\det\phi)_{t}=0$ by Theorem 3, thus we have
\begin{align}\label{Q17}
\det\phi_{\pm}(x,t;z)=\det E_{\pm}(z)=\gamma(z),\quad z\in\Sigma.
\end{align}
Assuming $\Sigma_{0}=\Sigma-\{\pm iq_{0}\}$, from Liouville's formula we get that $\phi_{\pm}$ are fundamental solutions of Lax pair \eqref{L1}, that means there exists a constant matrix $S(z)$ (it's independent of the variable $x$ and $t$) satisfying
\begin{align}\label{Q18}
 \phi_{+}(x,t;z)=\phi_{-}(x,t;z)S(z),\quad z\in\Sigma_{0},
\end{align}
which implies
\begin{align}\label{Q19}
\phi_{+,1}=s_{11}\phi_{-,1}+s_{21}\phi_{-,2},\quad
\phi_{+,2}=s_{12}\phi_{-,1}+s_{22}\phi_{-,2},
\end{align}
here $s_{ij}$ $(i, j=1, 2)$ are element of the matrix $S(z)$. From \eqref{Q13}, one has $\det S(z)=1$. Further the analyticity of the matrix $S(z)$ is discussed.

\noindent \textbf {Proposition 2.}
\emph{Assume $q-q_{\pm}\in L^{1}(\mathbb{R^{\pm}})$, afterwards the elements of $s_{11}$ and $s_{22}$ are analytically to $D_{+}$ and $D_{-}$, as well as continuously to $D_{+}\cup\Sigma_{0}$ and $D_{-}\cup\Sigma_{0}$, respectively. The off-diagonal elements of $S(z)$, although not analytical, continue to $\Sigma_{0}$.}
\begin{proof}
Resorting to \eqref{Q18}, one has
\begin{align}
&s_{11}(z)=\frac{Wr\left(\phi_{+,1},\phi_{-,2}\right)}{\gamma},\label{Wr-1}\quad
s_{22}(z)=\frac{Wr\left(\phi_{-,1},\phi_{+,2}\right)}{\gamma},\\
&s_{12}(z)=\frac{Wr\left(\phi_{+,2},\phi_{-,2}\right)}{\gamma},\quad
s_{21}(z)=\frac{Wr\left(\phi_{-,1},\phi_{+,1}\right)}{\gamma},
\end{align}
with $\gamma(z)=\det E_{\pm}(z)=1+q_{0}^{2}/z^{2}$. Based on Corollary 1., the Proposition is proved.
\end{proof}
Finally, we introduce the reflection coefficients that play an important role in the inverse problem given by
\begin{align}\label{Q21}
\rho(z)=s_{21}/s_{11},\quad \tilde{\rho}(z)=s_{12}/s_{22},\quad \forall z\in\Sigma.
\end{align}
\subsection{Symmetry of scattering matrix and Jost eigenfunction}
In this subsection, the symmetry of the scattering matrix and the eigenfunction will be obtained. What is important is that the scattering problem contains two symmetries, which are related to the values of Jost eigenfunctions  on each sheets of the Riemann surface, and the values of these Jost eigenfunctions will eventually affect the discrete spectral and residual conditions. The two symmetries are $(k,\lambda)\rightarrow(k^{*},\lambda^{*})$ and $(k,\lambda)\rightarrow(k,-\lambda)$ in $k$-plane map onto $z\rightarrow z^{*}$ and $z\rightarrow -q_{0}^{2}/z$ in $z$-plane by \eqref{Q7}.

\noindent \textbf {Proposition 3.} \emph{The symmetries for the Jost function $\phi_{\pm}\in\Sigma$ are presented for $z\in\Sigma$ as follows:}
\begin{subequations}
\begin{align}
\phi_{\pm}(z)&=-\sigma_{0}\phi_{\pm}^{*}(z^{*})\sigma_{0}, \label{symm-1}\\
\phi_{\pm}(z)&=\frac{i}{z}\phi_{\pm}(-\frac{q_{0}^{2}}{z})
\sigma_{3}Q_{\pm},\label{symm-2}
\end{align}
\end{subequations}
\emph{the above formulas can be expressed as the following column elements}
\begin{align}
\phi_{\pm,1}(z)&=\sigma_{0}\phi_{\pm,2}^{*}(z^{*}),\quad \qquad
\phi_{\pm,2}(z)=-\sigma_{0}\phi_{\pm,1}^{*}(z^{*}),\label{Sy-1}\\
\phi_{\pm,1}(z)&=(\frac{iq_{\pm}^{*}}{z})\phi_{\pm,2}(-\frac{iq_{0}^{2}}{z}),\quad
\phi_{\pm,2}(z)=(\frac{iq_{\pm}}{z})\phi_{\pm,1}(-\frac{iq_{0}^{2}}{z}).\label{Sy-2}
\end{align}

\noindent \textbf {Proposition 4.} \emph{The symmetries for the scattering matrix $S(z)$ is exhibited for $z\in\Sigma$ as follows:}
\begin{align}\label{Q22}
&S^{*}(z^{*})=-\sigma_{0}S(z)\sigma_{0},\\
&S(z)=(\sigma_{3}Q_{-})^{-1}S(-q_{0}^{2}/z)\sigma_{3}Q_{+}.
\end{align}

\noindent \textbf {Corollary 3.}
\emph{The relationship of the scattering coefficients and reflection coefficients using the above symmetries for $z\in\Sigma$}
\begin{align}
s_{22}(z)=s^{*}_{11}(z^{*}), \quad s_{12}(z)=-s^{*}_{21}(z^{*}),\label{chen-1}\\
s_{11}(z)=(q_{+}^{*}(z)/q_{-}^{*}(z))s_{22}(-q_{0}^{2}/z),\label{chen-2}\\
s_{12}(z)=(q_{+}(z)/q_{-}^{*}(z))s_{22}(-q_{0}^{2}/z),\\
\rho(z)=-\tilde{\rho}^{*}(z^{*})=(q^{*}_{-}/q_{-})\tilde{\rho}(-q_{0}^{2}/z).
\end{align}
\subsection{Discrete spectrum and residue condition}
The discrete spectrum of the scattering problem is composed of all values $k\in\mathbb{C}\setminus\Sigma$ meeting eigenfunctions exist in $L^{2}(\mathbb{R})$.
 Next we discuss that these discrete spectrum are the zeros of $s_{11}(z)$ and  $s_{22}(z)$ for $z\in\mathbb{D^{+}}$ and $z\in\mathbb{D^{-}}$, respectively. Assuming that $s_{11}(z)$ has $N$ simple zeros in $\mathbb{D^{+}}\cap\{z\in\mathbb{C}: Imz>0\}$ defined by $z_{n}$, $n=1, 2, \cdots, N$, namely, $s_{11}(z_{n})=0$ but $s'_{11}(z_{n})\neq0$, $n=1, 2, \cdots, N$. Recalling the symmetry properties \eqref{chen-1} and \eqref{chen-2}, we have $s_{22}(z_{n}^{*})=s_{22}(-q_{0}^{2}/z_{n})=s_{11}(-q_{0}^{2}/z_{n}^{*})=0$ if $s_{11}(z)=0$, which gives rise to the set of discrete spectrum
\begin{align}\label{Q23}
 Z^{d}=\left\{z_{n}, -\frac{q_{0}^{2}}{z_{n}^{*}},
  z_{n}^{*}, -\frac{q_{0}^{2}}{z_{n}}\right\}_{n=1}^{N},\quad s_{11}(z_{n})=0.
  \end{align}

In what follows, we investigate the residue conditions that will be required for inverse problem. When $z_{n}$ is a simple zero of $s_{11}(z)$ the relation can be derived by the first expression of \eqref{Wr-1}
\begin{align}\label{Q24}
\phi_{+,1}(z_{n})=b_{+}(z_{n})\phi_{-,2}(z_{n}),
\end{align}
which can write equivalently as
\begin{align}\label{Q25}
u_{+,1}(z_{n})=e^{-2i\theta(z_{n})}b_{+}(z_{n})u_{-,2}(z_{n}),
\end{align}
here $b_{+}(z_{n})$ is a normal constant. Thus,
\begin{align}\label{Q26}
\mathop{Res}_{z=z_{n}}\left[\frac{u_{+,1}(z)}{s_{11}(z)}\right]=
\frac{u_{+,1}(z_{n})}{s'_{11}(z_{n})}=\frac{b_{+}(z_{n})}{s'_{11}(z_{n})}
e^{-2i\theta(z_{n})}u_{-,2}(z_{n}).
\end{align}
Similarly, if $s_{22}(z_{n}^{*})=0$ and $s'_{22}(z_{n}^{*})\neq0$ for $\mathbb{D^{-}}\cap\{z\in\mathbb{C}: Imz<0\}$, one has
\begin{align}\label{Q27}
 u_{+,2}(z_{n}^{*})=b_{-}(z_{n}^{*})e^{2i\theta(z_{n}^{*})}u_{-,1}(z_{n}^{*}),
\end{align}
  as well as the following residue conditions from the second expression of \eqref{Wr-1}
\begin{align}\label{Q28}
\mathop{Res}_{z=z_{n}^{*}}\left[\frac{u_{+,2}(z)}{s_{22}(z)}\right]=
\frac{u_{+,2}(z_{n}^{*})}{s'_{22}(z_{n}^{*})}=
\frac{b_{-}(z_{n}^{*})}{s'_{22}(z_{n}^{*})}
e^{2i\theta(z_{n^{*}})}u_{-,1}(z_{n}^{*}).
\end{align}
For convenience, let's take a brief note
\begin{align}
 C_{+}[z_{n}]&=\frac{b_{+}(z_{n})}{s'_{11}(z_{n})},\\
C_{-}[z_{n}]&=\frac{b_{-}(z_{n}^{*})}{s'_{22}(z_{n}^{*})}.
\end{align}
\noindent \textbf {Corollary 4.}
\emph{The coefficients of the residue conditions are of the relation}
\begin{align}\label{Q29}
C_{+}[z_{n}]=-C_{-}^{*}[z_{n}^{*}],\quad C_{+}[z_{n}]=\frac{z_{n}^{2}}{q_{-}^{2}}
C_{-}\left[-\frac{q_{0}^{2}}{z_{n}}\right].
\end{align}
\noindent \textbf {Corollary 4.1}
\emph{From the symmetry of scattering matrix and Jost function, the following relations can be further deduced.}
\begin{align}
C_{+}[z_{n}]=-C_{-}^{*}[z_{n}^{*}]=\frac{z_{n}^{2}}{q_{-}^{2}}
C_{-}\left[-\frac{q_{0}^{2}}{z_{n}}\right]=-\frac{z_{n}^{2}}{q_{-}^{2}}
C_{+}^{*}\left[-\frac{q_{0}^{2}}{z_{n}^{*}}\right].
\end{align}

\subsection{Asymptotic analysis}
The asymptotic property of the Jost function and  scattering data can be used to establish a suitable RH problem, and the potential formula can be reconstructed by solving the RH problem. Thus we next consider the asymptotic behavior in the case that $z$ tends to infinity and zero, respectively. Consider the Neumann series
\begin{align}
u_{\pm}(x,t;z)=\sum_{n=0}^\infty u_{\pm}^{[n]}(x,t;z),
\end{align}
with
\begin{subequations}
\begin{align}
&u_{\pm}^{[0]}(x,t;z)=E_{\pm}(z),\\
u_{\pm}^{[n+1]}(x,t;z)=\int_{\pm\infty}^{x}E_{\pm}(z)&e^{i\lambda(x-y)\sigma_{3}}\left[
E_{\pm}^{-1}(z)\Delta Q_{\pm}(y,t)u_{\pm}^{[n]}(y,t;z)\right]e^{-i\lambda(x-y)\sigma_{3}}\, dy.
\end{align}
\end{subequations}
Similar to \cite{Biondini}, combined with \cite{jianjin}, we can get
\begin{align}
\begin{split}
u_{\pm}^{[n+1],d}(x,t;z)=\frac{1}{1+(q_{0}/z)^{2}}\int_{\pm\infty}^{x}\left(
\Delta Q_{\pm}(y,t)u_{\pm}^{[n],o}(y,t;z)-\frac{i\sigma_{3}Q_{\pm}(y,t)}{z}
\Delta Q_{\pm}(y,t)u_{\pm}^{[n],d}(y,t;z)\right)\, dy \\+
\frac{i\sigma_{3}Q_{\pm}(y,t)}{z(1+(q_{0}/z)^{2})}\int_{\pm\infty}^{x}e^{i\lambda(x-y)\hat{\sigma}_{3}}
\left(\Delta Q_{\pm}(y,t)u_{\pm}^{[n],d}(y,t;z)-\frac{i\sigma_{3}Q_{\pm}(y,t)}{z}
\Delta Q_{\pm}(y,t)u_{\pm}^{[n],o}(y,t;z)\right)\ dy \\
=\left\{
\begin{aligned}
&O\left(u_{\pm}^{[n],o}(y,t;z)\right)+O\left(\frac{u_{\pm}^{[n],d}(y,t;z)}{z}\right)
+O\left(\frac{u_{\pm}^{[n],d}(y,t;z)}{z^{2}}\right)+
O\left(\frac{u_{\pm}^{[n],o}(y,t;z)}{z^{3}}\right), \quad &z\rightarrow\infty,\\
&O\left(z^{2}u_{\pm}^{[n],o}(y,t;z)\right)+O\left(zu_{\pm}^{[n],d}(y,t;z)\right)+
O\left(z^{2}u_{\pm}^{[n],d}(y,t;z)\right)+O\left(zu_{\pm}^{[n],o}(y,t;z)\right), \quad &z\rightarrow0,
\end{aligned}
\right.
\end{split}
\end{align}
\begin{align}
\begin{split}
u_{\pm}^{[n+1],o}(x,t;z)=\frac{i\sigma_{3}Q_{\pm}(y,t)}{z(1+(q_{0}/z)^{2})}
\int_{\pm\infty}^{x}\left(
\Delta Q_{\pm}(y,t)u_{\pm}^{[n],o}(y,t;z)-\frac{i\sigma_{3}Q_{\pm}(y,t)}{z}
\Delta Q_{\pm}(y,t)u_{\pm}^{[n],d}(y,t;z)\right)\, dy \\
+\frac{1}{1+(q_{0}/z)^{2}}
\int_{\pm\infty}^{x}e^{i\lambda(x-y)\hat{\sigma}_{3}}
\left(\Delta Q_{\pm}(y,t)u_{\pm}^{[n],d}(y,t;z)-\frac{i\sigma_{3}Q_{\pm}(y,t)}{z}
\Delta Q_{\pm}(y,t)u_{\pm}^{[n],o}(y,t;z)\right)\ dy \\
=\left\{
\begin{aligned}
&O\left(\frac{u_{\pm}^{[n],o}(y,t;z)}{z}\right)+O\left(\frac{u_{\pm}^{[n],d}(y,t;z)}{z^{2}}\right)
+O\left(\frac{u_{\pm}^{[n],d}(y,t;z)}{z}\right)+
O\left(\frac{u_{\pm}^{[n],o}(y,t;z)}{z^{2}}\right), \quad &z\rightarrow\infty,\\
&O\left(zu_{\pm}^{[n],o}(y,t;z)\right)+O\left(u_{\pm}^{[n],d}(y,t;z)\right)+
O\left(z^{3}u_{\pm}^{[n],d}(y,t;z)\right)+O\left(z^{2}u_{\pm}^{[n],o}(y,t;z)\right), \quad &z\rightarrow0,
\end{aligned}
\right.
\end{split}
\end{align}
here the superscript $o$ and $d$ of $u_{\pm}$ represent the diagonal and off-diagonal of $u_{\pm}$, respectively. At this point, we also temporarily regard time as a dumb variable. From the above formula, the realtionship can be concluded as follow
\begin{align}
\begin{split}
u_{\pm}^{[0],d}(x,t;z)=O(1),\quad u_{\pm}^{[0],o}(x,t;z)=O(1/z)\quad z\rightarrow\infty,\\
u_{\pm}^{[0],d}(x,t;z)=O(1),\quad u_{\pm}^{[0],o}(x,t;z)=O(1/z)\quad z\rightarrow0,
\end{split}
\end{align}
furthermore, for all $n\in\mathbb{N}$, one has
\begin{align}
\left\{
\begin{aligned}
u_{\pm}^{[2n],d}(x,t;z)&=O(1/z^{n}), \quad\quad u_{\pm}^{[2n],o}(x,t;z)=O(1/z^{n+1}),\\
u_{\pm}^{[2n+1],d}(x,t;z)&=O(1/z^{n+1}),\quad u_{\pm}^{[2n+1],o}(x,t;z)=O(1/z^{n+1}),
\end{aligned}
\right. \quad z\rightarrow\infty,\\
\left\{\begin{aligned}
u_{\pm}^{[2n],d}(x,t;z)=O(z^{n}),\quad u_{\pm}^{[2n],o}(x,t;z)=O(z^{n-1}),\\
u_{\pm}^{[2n+1],d}(x,t;z)=O(z^{n}),\quad u_{\pm}^{[2n+1],d}(x,t;z)=O(z^{n}),
\end{aligned}
\right.\quad z\rightarrow0.
\end{align}
\noindent \textbf {Corollary 5.}
\emph{The asymptotic behavior of the Jost function can be defined as in the appropriate regions of the $z$-plane}
\begin{align}
u_{\pm}(x,t;z)=
\left\{
\begin{aligned}
&I+O(1/z),\quad\quad &z\rightarrow\infty,\\
&\frac{i}{z}\sigma_{3}Q_{\pm}+O(1),\quad &z\rightarrow0.
\end{aligned}
\right.
\end{align}
\begin{proof}
In general, we prove it only in one case $z\rightarrow\infty$, and in another case $z\rightarrow0$  it can be similar. Consider the series
\begin{align}\label{se-1}
u_{\pm}=u_{\pm}^{(0)}+u_{\pm}^{(1)}/z+o(1/z), z\rightarrow\infty.
\end{align}
Substituting \eqref{se-1} into \eqref{Q15}  and then comparing the powers of the function, one has the matrix $u_{\pm}^{(0)}$ is a diagonal matrix and is independent of variables $x$ and $t$, thus it is possible to exchange the order of integrals for the function $u_{\pm}(z)$ tending to infinite limit in $z$ and $x$. After analysis, we know that
\begin{align}
\lim_{z\rightarrow\infty}\lim_{x\rightarrow\pm\infty}u_{\pm}=
\lim_{x\rightarrow\pm\infty}\lim_{z\rightarrow\infty}(u_{\pm}^{(0)}+o(1/z))=u_{\pm}^{0},
\end{align}
and \eqref{Q14} implies $u_{\pm}^{(0)}=\mathbb{I}$. It is worth noting that function $E_{\pm}$ is also a function of $z$. So, we're done proving the Corollary.
\end{proof}

\noindent \textbf {Corollary 6.}
\emph{The asymptotic behavior of the scattering matrix can be derived by in the appropriate regions of the $z$-plane}
\begin{align}
S(z)&=I+O(1/z),  \qquad \quad\qquad\qquad z\rightarrow\infty,\label{S-1}\\
S(z)&=diag(q_{-}/q_{+},q_{+}/q_{-})+O(z),\quad z\rightarrow0.\label{S-2}
\end{align}
\begin{proof}
Based on Corollary 4 and the first expression of \eqref{Wr-1}, one has
\begin{align}
s_{11}(z)&=\frac{Wr\left(\phi_{+,1},\phi_{-,2}\right)}{\gamma}=
\frac{Wr\left(u_{+,1},u_{-,2}\right)}{1+q_{0}^{2}/z^{2}}\\&=
\left\{\begin{aligned}
&\frac{\det\left(
        \begin{array}{cc}
          1+O(1/z) & O(1/z) \\
          O(1/z) & 1+O(1/z) \\
        \end{array}
      \right)
}{1+Q(1/z^{2})}=1+O(1/z), \quad z\rightarrow\infty,\\
&\frac{\det\left(
        \begin{array}{cc}
          O(1) & (i/z)q_{-} \\
          (i/z)q_{+} & O(1) \\
        \end{array}
      \right)
}{q_{0}^{2}+Q(z^{2})}z^{2}=\frac{q_{-}}{q_{+}}+O(z),
\quad\quad\quad\quad z\rightarrow0.
\end{aligned}
\right.
\end{align}
\end{proof}
The element asymptotic property of residual matrix can be obtained similarly. Now the Corollary is completed.
\section{Inverse scattering problem with the simple poles}
\subsection{Generalized Riemann-Hilbert problem}
In order to give a generalized Riemann-Hilbert problem, we need to redefine sectionally analytic functions according to the analyticity of scattering matrix and eigenfunction so that they can be analyzed in $D^{+}$ and $D^{-}$, respectively. The following propositions are given in detail.

\noindent \textbf {Proposition 5.}
\emph{We define the sectionally  meromorphic matrices}
\begin{align}\label{Matr}
M(x,t;z)=\left\{\begin{aligned}
&M^{+}(x,t;z)=\left(\frac{u_{+,1}(x,t;z)}{s_{11}(z)},u_{-,2}(x,t;z)\right), \quad z\in D^{+},\\
&M^{-}(x,t;z)=\left(u_{-,1}(x,t;z),\frac{u_{+,2}(x,t;z)}{s_{22}(z)}\right), \quad z\in D^{-}.
\end{aligned}\right.
\end{align}

From the above proposition, the multiplicative matrix RH problem is constructed as follow

\noindent \textbf {Proposition 6.}
\begin{align}\label{RHP}
M^{-}(x,t;z)=M^{+}(x,t;z)(\mathbb{I}-G(x,t;z)),
\end{align}
\emph{with the jump matrix}
\begin{align*}
G(x,t;z)=e^{i\theta(z)\hat{\sigma}_{3}}\left(
\begin{array}{ccc}
  0 & -\tilde{\rho}(z) \\
  \rho(z) & \rho(z)\tilde{\rho}(z)
\end{array} \right)
\end{align*}
\emph{and the asymptotic behavior}
\begin{align}\label{jianjin}
M^{\pm}(x,t;z)=\left\{
\begin{aligned}
\mathbb{I}+O(1/z), \quad z\rightarrow\infty,\\
(i/z)\sigma_{3}Q_{-}+O(1), z\rightarrow0.
\end{aligned}
\right.
\end{align}
\emph{For the convenience of solving Riemann Hilbert problem later, we introduce}
\begin{align}\label{F3}
\xi_{n}=\left\{\begin{aligned}
&z_{n}, \quad\quad n=1, 2, \cdots,N,\\
-&\frac{q_{0}^{2}}{z_{n-N}^{*}}, \quad n=N+1, N+2, \cdots, 2N,
\end{aligned}\right.
\end{align}
\emph{here $z_{n}$ is the zero of $s_{11}(z)$ and $\hat{\xi}_{n}=-q_{0}^{2}/\xi_{n}$.}

\noindent \textbf {Theorem 4.}
\emph{The solutions of the matrix RHP \eqref{RHP} can be written as}
\begin{align}\label{jie-M}
\begin{split}
 M(x,t;z)=\mathbb{I}+\frac{i}{z}\sigma_{3}Q_{-}&+\sum_{n=1}^{2N}\frac
{Res_{z=\xi_{n}}M(z)}{z-\xi_{n}}+\sum_{n=1}^{2N}\frac
{Res_{z=\hat{\xi}_{n}}M(z)}{z-\hat{\xi}_{n}}\\
&+\frac{1}{2i\pi}\int_{\Sigma}\frac{M(x,t;\zeta)G(x,t;\zeta)}{\zeta-z}\,d\zeta,\quad
z\in\mathbb{C}\setminus\Sigma,
\end{split}
\end{align}
\emph{where $\int_{\Sigma}$ means the counter shown in Fig. 1.}

\begin{proof}
The matrix RHP \eqref{RHP}, for given the condition as Proposition 6, can be regularized in terms of subtracting out the asymptotic behaviors and pole contributions, then one has
\begin{align}\label{PL-1}
\begin{split}
 M^{-}(x,t;z)-&\mathbb{I}-\frac{i}{z}\sigma_{3}Q_{-}-\sum_{n=1}^{2N}\frac
{Res_{z=\hat{\xi}_{n}}M^{-}(z)}{z-\hat{\xi}_{n}}-\sum_{n=1}^{2N}\frac
{Res_{z=\xi_{n}}M^{+}(z)}{z-\xi_{n}}\\&=
M^{+}(x,t;z)-\mathbb{I}-\frac{i}{z}\sigma_{3}Q_{-}-\sum_{n=1}^{2N}\left[\frac
{Res_{z=\xi_{n}}M^{+}(z)}{z-\xi_{n}}-\frac
{Res_{z=\hat{\xi}_{n}}M^{-}(z)}{z-\hat{\xi}_{n}}\right]-M^{+}(z)G(z),
\end{split}
\end{align}
we know that left hand of \eqref{PL-1} is analytic in $D^{-}$, meanwhile, the first four terms for right hand of \eqref{PL-1} is analytic in $D^{+}$. Then \eqref{PL-1} can be solved by Plemelj's formulate. Introduce the Cauchy projectors $P_{\pm}$ over $\Sigma$ by
\begin{align}
P_{\pm}[f](z)=\frac{1}{2i\pi}\int_{\Sigma}\frac{f(\zeta)}{\zeta-(z\pm i0)},
\end{align}
here $z\pm i0$ denotes that the limit is taken from the left/right for $z\in\Sigma$. According to Cauchy operator Properties, when $f^{\pm}$ are analytic in $D^{\pm}$, respectively, one has $P^{\pm}f^{\pm}=\pm f^{\pm}$ and $P^{+}f^{-}=P^{-}f^{+}=0$
Applying the Cauchy operator to \eqref{PL-1}, one can get the \eqref{jie-M}. Thus the theorem is proved.
\end{proof}
\subsection{Reconstruct the formula for potential}
The solution of the RH problem, for a closed system, can be presented by the expression of residue condition in \eqref{jie-M}. From the expression \eqref{Matr}, only the first column of $M$ has a pole at point $z=z_{n}$ and $z=-q_{0}^{2}/z_{n}^{*}$, as well as only the second column has a pole at point $z=z_{n}^{*}$ and $z=-q_{0}^{2}/z_{n}$, namely
\begin{align}\label{liu}
\begin{split}
\mathop{Res}_{z=\xi_{n}}M^{+}&=(C_{+}[\xi_{n}]e^{-2i\theta(\xi_{n})}u_{-,2}(\xi_{n}),0),\quad n=1, 2, \cdots, 2N,\\
\mathop{Res}_{z=\hat{\xi}_{n}}M^{-}&=(0,C_{-}[\hat{\xi}_{n}]e^{2i\theta(\hat{\xi}_{n})}
u_{-,1}(\hat{\xi}_{n})),\quad n=1, 2, \cdots, 2N.
\end{split}
\end{align}
Further for $z=\xi_{s}$ $(s=1,2,\cdots,2N)$
\begin{align}\label{jie-1}
u_{-,2}(x,t;\xi_{s})=\left(\begin{array}{cc}
                       iq_{-}/\xi_{s} \\
                        1
                     \end{array}\right)
+\sum_{k=1}^{2N}
\frac{C_{-}[\hat{\xi}_{n}]e^{2i\theta(\hat{\xi}_{n})}}{\xi_{s}-\hat{\xi}_{n}}
u_{-,1}(x,t;\hat{\xi}_{n})
+\frac{1}{2i\pi}\int_{\Sigma}\frac{(M^{+}G)_{2}(x,t;\xi)}{\xi-\xi_{s}}\,d\xi.
\end{align}
Then \eqref{Sy-2} gives rise to
\begin{align}\label{F1}
u_{-,2}(x,t;\xi_{s})=iq_{-}/\xi_{s}u_{-,1}(x,t;\hat{\xi_{s}}),\quad s=1,2,\cdots,2N.
\end{align}
Substituting the \eqref{F1} into \eqref{jie-1}, one has
\begin{align}\label{F2}
\sum_{n=1}^{2N}\left(\frac{C_{-}[\hat{\xi}_{n}]e^{2i\theta(\hat{\xi}_{n})}}
{\xi_{s}-\hat{\xi}_{n}}-\frac{iq_{-}}{\xi_{s}}\delta_{sn}\right)
u_{-,1}(x,t;\hat{\xi}_{n})+\left(\begin{array}{cc}
                       iq_{-}/\xi_{s} \\
                        1
                     \end{array}\right)+
  \frac{1}{2i\pi}\int_{\Sigma}\frac{(M^{+}G)_{2}(x,t;\xi)}{\xi-\xi_{s}}\,d\xi=0,
\end{align}
with $\delta_{sn}$ is the Kronecker delta function. From the equation \eqref{F2}   we know that it contains $2N$ equations with $2N$ unknowns, and then $2N$ solutions  $u_{-,1}(x,t;\hat{\xi_{s}})$  $(s=1,2,\cdots,2N)$ can be generated, as well as $u_{-,2}(x,t;\xi_{s})$ $(s=1,2,\cdots,2N)$ can be derived via \eqref{F1}. Thus the form of function $M(x,t;z)$  can be obtained based on scattering data by introducing the expression $u_{-,1}(x,t;\hat{\xi_{s}})$ and $u_{-,2}(x,t;\xi_{s})$ into \eqref{liu}  and then substituting \eqref{liu} into \eqref{jie-M}.

Finally, we need to accomplish the task of reconstructing potential function.

\noindent \textbf {Theorem 5.}
\emph{The potential function with simple poles in modified Landau-Lifshitz
equation with NZBCs is derived by}
\begin{align}\label{JIE}
q(x,t)=q_{-}-i\sum_{n=1}^{2N}C_{-}[\hat{\xi}_{n}]e^{2i\theta(\hat{\xi}_{n})}
u_{-,1,1}(x,t;\hat{\xi}_{n})+\frac{1}{2\pi}\int_{\Sigma}(M^{+}G)_{12}(x,t;\xi)\,d\xi,
\end{align}
\emph{here $\xi_{n}$ is determined by \eqref{F3}, and $u_{-,1,1}(x,t;\hat{\xi}_{n})$ is defined by}
\begin{align*}
\sum_{n=1}^{2N}\left(\frac{C_{-}[\hat{\xi}_{n}]e^{2i\theta(\hat{\xi}_{n})}}
{\xi_{s}-\hat{\xi}_{n}}-\frac{iq_{-}}{\xi_{s}}\delta_{sn}\right)
u_{-,1}(x,t;\hat{\xi}_{n})+ \frac{iq_{-}}{\xi_{s}} +
  \frac{1}{2i\pi}\int_{\Sigma}\frac{(M^{+}G)_{1,2}(x,t;\xi)}{\xi-\xi_{s}}\,d\xi=0.
  \end{align*}

\begin{proof}
Now $M(x,t;z)e^{i\theta(x,t;z)\sigma_{3}}$ is the solution of \eqref{QQ1}, which implies
\begin{align}\label{jie-3}
M_{x}(x,t;z)+M(x,t;z)\left(\frac{1}{2}i\sigma_{3}z+\frac{1}{2z}iq_{0}^{2}
\sigma_{3}\right)=\left(\frac{1}{2}i\sigma_{3}z-\frac{1}{2z}iq_{0}^{2}
\sigma_{3}+Q\right)M(x,t;z).
\end{align}
Making Taylor expansion of functions $M$
\begin{align}\label{Taylor}
M(x,t;z)=\mathbb{I}+\frac{1}{z}M^{(1)}(x,t;z)+O(1/z^{2}), \quad z\rightarrow\infty,
\end{align}
one knows that the expression of $M^{(1)}(x,t;z)$ via \eqref{jie-M} and \eqref{liu}, i.e.,
\begin{align}\label{jie-4}
\begin{split}
M^{(1)}(x,t;z)=i\sigma_{3}q_{-}-\frac{1}{2\pi i}\int_{\Sigma}M^{+}(x,t;\zeta)G(x,t;\zeta)\,d\zeta\\+
\sum_{n=1}^{2N}(C_{+}[\xi_{n}]e^{-2i\theta(\xi_{n})}u_{-,2}(\xi_{n}),
C_{-}[\hat{\xi}_{n}]e^{2i\theta(\hat{\xi}_{n})}u_{-,2}(\hat{\xi}_{n})).
\end{split}
\end{align}
By introducing \eqref{Taylor} into \eqref{jie-3} and comparing the coefficients of function $z^{0}$, we can get the result
\begin{align}
\frac{i}{2}[M^{(1)},\sigma_{3}]=Q,
\end{align}
which means
\begin{align}\label{jie-5}
q(x,t)=-i(M^{(1)})_{12},
\end{align}
therefore we can conclude the theorem from the \eqref{jie-4} and \eqref{jie-5}.
\end{proof}
\subsection{Trace formulate and theta condition}
In this subsection, the scattering coefficients $s_{11}(z)$ and $s_{22}(z)$ are expressed by reflection coefficients $\rho(z)$ and $\tilde{\rho}(z)$ and discrete spectral points $Z^{d}$, i.e., the so-called trace formula. At the same time, the phase difference of boundary conditions $q_{+}$ and $q_{-}$, i.e.,
 theta     conditions \cite{LD}, will be derived. Proposition 1 implies that $s_{11}(z)$ and $s_{22}(z)$ are analytic in $D^{+}$ and $D^{-}$, respectively.
Further considering the following function
\begin{align}\label{trace}
\vartheta^{+}(z)=s_{11}(z)\prod_{n=1}^{2N}\frac{(z-z_{n}^{*})(z+q_{0}^{2}/z_{n})}
{(z-z_{n})(z+q_{0}^{2}/z_{n}^{*})},\quad
\vartheta^{-}(z)=s_{22}(z)\prod_{n=1}^{2N}\frac{(z-z_{n})(z+q_{0}^{2}/z_{n}^{*})}
{(z-z_{n}^{*})(z+q_{0}^{2}/z_{n})},
\end{align}
we can get $\vartheta^{+}(z)$ and $\vartheta^{-}(z)$ are analytic and no zeros in $D^{+}$ and $D^{-}$. The equation \eqref{S-1} means $\vartheta^{\pm}(z)\rightarrow 1$ as $z\rightarrow\infty$. Taking determinants of \eqref{Q18}, one has
\begin{align}\label{T14}
\det S(z)=s_{11}(z)s_{22}(z)-s_{12}(z)s_{21}(z)=1, \quad z\in\Sigma,
\end{align}
which leads to
\begin{align}\label{det}
\vartheta^{+}(z)\vartheta^{-}(z)=\frac{1}{1-\rho(z)\tilde{\rho}(z)},\quad z\in\Sigma.
\end{align}
Taking logarithms of \eqref{det}, we have
\begin{align}
\log\vartheta^{+}(z)+\log\vartheta^{-}(z)=
-\log[1-\rho(z)\tilde{\rho}(z)], \quad z\in\Sigma.
\end{align}
Furthermore
\begin{align}\label{log}
\log\vartheta^{\pm}(z)=\mp\frac{1}{2\pi i}\int_{\Sigma}
\frac{\log[1-\rho(\zeta)\tilde{\rho}(\zeta)]}{\zeta-z}\,d\zeta,\quad z\in D^{\pm},
\end{align}
with the help of Plemelj's formulae and Cauchy projectors.

By bring \eqref{log} into \eqref{trace}, the trace formulate can be given by
\begin{align}
s_{11}(z)&=exp\left(-\frac{1}{2\pi}\frac{\log[1-\rho(\zeta)\tilde{\rho}(\zeta)]}{\zeta-z}
\,d\zeta\right)\prod_{n=1}^{2N}\frac{(z-z_{n})(z+q_{0}^{2}/z_{n}^{*})}
{(z-z_{n}^{*})(z+q_{0}^{2}/z_{n})},\label{s11}\\
s_{22}(z)&=exp\left(\frac{1}{2\pi}\frac{\log[1-\rho(\zeta)\tilde{\rho}(\zeta)]}{\zeta-z}
\,d\zeta\right)\prod_{n=1}^{2N}\frac{(z-z_{n}^{*})(z+q_{0}^{2}/z_{n})}
{(z-z_{n})(z+q_{0}^{2}/z_{n}^{*})}.\label{s22}
\end{align}

We now discuss the asymptotic phase of the boundary conditions $q_{+}$ and $q_{-}$. The equation \eqref{S-2} shows $s_{11}(z)\rightarrow q_{-}/q_{+}$. Note that
\begin{align}
\prod_{n=1}^{2N}\frac{(z-z_{n}^{*})(z+q_{0}^{2}/z_{n})}
{(z-z_{n})(z+q_{0}^{2}/z_{n}^{*})}\rightarrow 1, \quad for \quad z\rightarrow0.
\end{align}
Thus from \eqref{s11}, we can get
\begin{align}
q_{-}/q_{+}=exp\left(\frac{i}{2\pi}\int_{\Sigma}
\frac{\log[1-\rho(\zeta)\tilde{\rho}(\zeta)]}{\zeta}\,d\zeta\right),\quad for \quad z\rightarrow 0,
\end{align}
which can be converted to the theta condition
\begin{align}
\arg\frac{q_{-}}{q_{+}}=\frac{1}{2\pi}\int_{\Sigma}
\frac{\log[1-\rho(\zeta)\tilde{\rho}(\zeta)]}{\zeta}\,d\zeta+4\sum_{n=1}^{N}\arg z_{n}.
\end{align}
If we consider the case of non-scattering potential coefficients, that is, the function $\rho(z)=0$, the phase difference is defined by
\begin{align}\label{F5}
\arg\frac{q_{-}}{q_{+}}=\arg q_{-}-\arg q_{+}=4\sum_{n=1}^{N}\arg z_{n}.
\end{align}
\subsection{Reflection-less potentials }
In this section, we will give the special solution of the equation \eqref{QQ1} without scattering coefficient ($\rho(z)=\tilde{\rho}(z)=0$), which means that there is no jump matrix (i,e,. $J=0$) along the continuous spectrum from $M^{+}(x,t;z)$ to $M^{-}(x,t;z)$. Thus \eqref{F2} can reduce to
\begin{align}\label{F4}
 \sum_{n=1}^{2N}\left(\frac{C_{-}[\hat{\xi}_{n}]e^{2i\theta(\hat{\xi}_{n})}}
{\xi_{s}-\hat{\xi}_{n}}-\frac{iq_{-}}{\xi_{s}}\delta_{sn}\right)
u_{-,1}(x,t;\hat{\xi}_{n})+iq_{-}/\xi_{s}=0.
\end{align}
Then the closed system can be solved algebraically to obtain soliton solutions.

\noindent \textbf {Proposition 7.}
\emph{The non-zero boundary value form solution of the equation \eqref{QQ1}  without scattering coefficient can be expressed as}
\begin{align}\label{Jie-Q}
q(x,t)=q_{-}+i\frac{\det\tilde{M}}{\det M},
\end{align}
\emph{with $\det\tilde{M}=\left(\begin{array}{cc}
                      M & \nu \\
                      \omega^{T} & 0
                    \end{array}\right)$,
here $\omega=(\omega_{j})_{(2N)\times1}$, $\nu=(\nu_{j})_{(2N)\times1}$, $M=(m_{sj})_{(2N)\times(2N)}$, and $Y=(y_{n})_{(2N)\times1}$  with $\omega_{k}=C_{-}[\hat{\xi_{j}}]e^{2i\theta(x,t;\hat{\xi_{j}})}$, $\nu_{j}=-\frac{iq_{-}}{\xi_{j}}$, $m_{sj}=\frac{\omega_{j}}{\xi_{s-\hat{\xi}_{j}}}$, and $y_{n}=u_{-,1,1}(x,t;\hat{\xi}_{n})$  for $n=1, 2, \cdots, 2N$.}

\section{Soliton solutions}

Note that for some fixed parameters $\delta_{1}$, $\delta_{2}$, $z_{n}$, the reside condition $C_{+}[z_{n}]$ and $N$, the explicit expression of the formula \eqref{Jie-Q} can be presented via the Maple. Generally speaking, the forms of these solutions are very complex. For the sake of the coherence of the paper, we omit the formal solutions of the equations \eqref{Jie-Q} under specific parameters, and only give the propagation behavior of the solutions under specific parameters and a brief analysis of these images.

\noindent \textbf {Case 1.}
When $N=1$, $z_{1}=3i/2$, the asymptotic phase difference is $2\pi$ from \eqref{F5}, which means that there is no phase transition. We will get the breather solution of
 the modified Landau-Lifshitz equation with simple-pole.

{\rotatebox{0}{\includegraphics[width=2.75cm,height=2.5cm,angle=0]{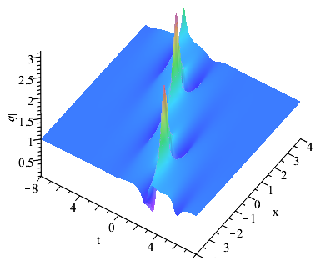}}}
{\rotatebox{0}{\includegraphics[width=2.75cm,height=2.5cm,angle=0]{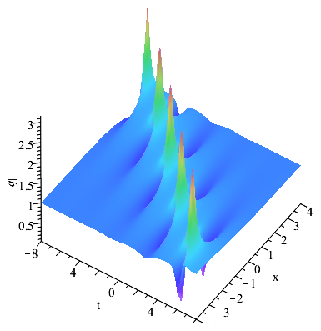}}}
{\rotatebox{0}{\includegraphics[width=2.75cm,height=2.5cm,angle=0]{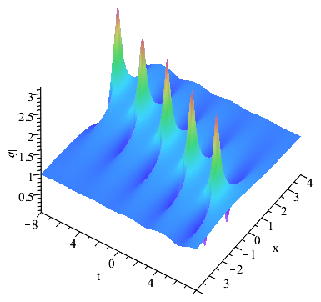}}}
{\rotatebox{0}{\includegraphics[width=2.75cm,height=2.5cm,angle=0]{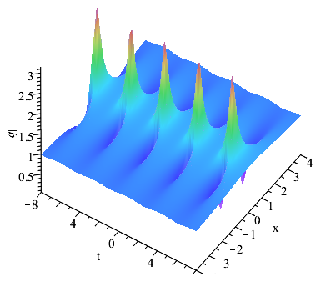}}}

 $\qquad\quad(\textbf{a})\qquad \ \qquad\quad\qquad(\textbf{b})
\ \qquad\quad\quad\quad\qquad(\textbf{c})\qquad\qquad\quad\qquad(\textbf{d})
\ \qquad$\\
\noindent { \small \textbf{Figure 2.} (Color online) The breather wave solutions  with the fixed parameters $N=1$, $z_{1}=3i/2$, $C_{+}[Z_{1}]=1$ and $q_{-}=1$. $\textbf{(a)}$: the breather wave solution with the $\delta_{2}=1$; $\textbf{(b)}$: the breather wave solution with the $\delta_{2}=0.5$; $\textbf{(c)}$: the breather wave solution with the $\delta_{2}=0.2$; $\textbf{(d)}$: the breather wave solution with the $\delta_{2}=0$;

It can be seen from the Figure $2$ that with the decrease of the parameters $\delta_{2}$, the propagation of the solution is gradually parallel to the time axis. When the parameters are zero, the solution of the equation is completely parallel to the time axis, which is called the stationary breathing solution or the time-periodic breather solution. Because the solution has good behavior without parameter influence, we should consider it further. How does the given boundary condition affect the propagation behavior of the solution?

{\rotatebox{0}{\includegraphics[width=2.75cm,height=2.5cm,angle=0]{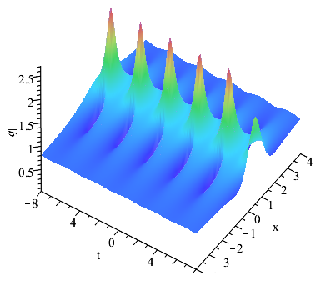}}}
{\rotatebox{0}{\includegraphics[width=2.75cm,height=2.5cm,angle=0]{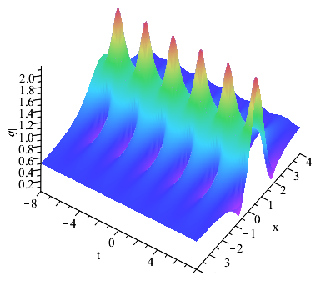}}}
{\rotatebox{0}{\includegraphics[width=2.75cm,height=2.5cm,angle=0]{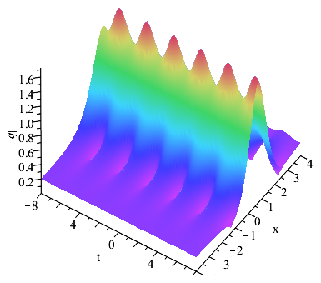}}}
{\rotatebox{0}{\includegraphics[width=2.75cm,height=2.5cm,angle=0]{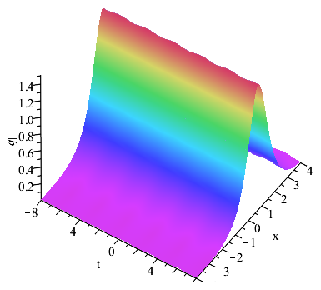}}}

 $\qquad\qquad(\textbf{a})\qquad \ \qquad\qquad\qquad(\textbf{b})
\ \qquad\quad\quad\quad\qquad(\textbf{c})\qquad\qquad\qquad\qquad(\textbf{d})
\ \qquad$\\
\noindent { \small \textbf{Figure 3.} (Color online)  The breather wave solutions  with the fixed parameters $N=1$, $z_{1}=3i/2$, $C_{+}[z_{1}]=1$. $\textbf{(a)}$: the breather wave solution with the $q_{-}=0.8$; $\textbf{(b)}$: the breather wave solution with the $q_{-}=0.5$; $\textbf{(c)}$: the breather wave solution with the $q_{-}=0.2$; $\textbf{(d)}$: the bright soliton solution with the $q_{-}=0.01$;

From the above Figure $3(a)-3(d)$, we can see that the periodic behavior of the solution gradually moves up with the decrease of the boundary condition, that is, the periodic behavior of the simple-pole breather wave solution occurs at the top of the solution, and when the boundary condition tends to zero, the breather wave solution  will eventually be transformed into a bright soliton solution.

In case $1$, the discrete spectrum we selected is a pure imaginary eigenvalue. Next, we consider the case where the spectral point has a real part.

\noindent \textbf {Case 2.}
For $N=1$, $q_{-}=1$, and $C_{+}[z_{1}]$,  we choose two complex eigenvalue $z_{1}=2e^{i\pi/3}$, $z_{1}=2e^{2i\pi/3}$, and the corresponding asymptotic phase difference are $4\pi/3$ and $8\pi/3$, respectively.

{\rotatebox{0}{\includegraphics[width=2.85cm,height=2.8cm,angle=0]{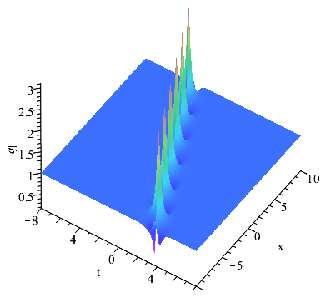}}}
{\rotatebox{0}{\includegraphics[width=2.35cm,height=2.15cm,angle=0]{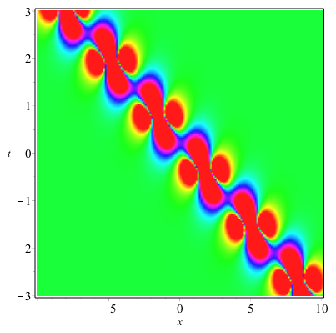}}}
{\rotatebox{0}{\includegraphics[width=2.85cm,height=2.8cm,angle=0]{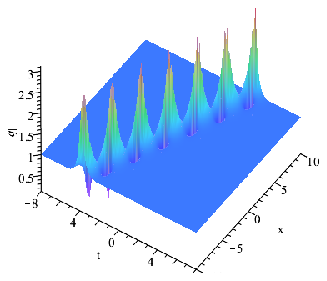}}}
{\rotatebox{0}{\includegraphics[width=2.35cm,height=2.15cm,angle=0]{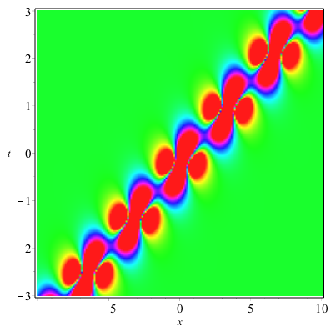}}}

 $\qquad\qquad(\textbf{a})\quad \ \qquad\qquad\qquad(\textbf{b})
\ \qquad\quad\quad\quad\qquad(\textbf{c})\qquad\qquad\qquad\qquad(\textbf{d})
\ \qquad$\\
\noindent { \small \textbf{Figure 4.} (Color online)  The breather wave   solutions  with the fixed parameters $N=1$,  $C_{+}[z_{1}]=1$ and $\delta_{2}=0$. $\textbf{(a)}$: the breather wave solution with the $q_{-}=1$, $z_{1}=2e^{i\pi/3}$; $\textbf{(b)}$: shows the density; Fig. $\textbf{(c)}$: the breather wave solution with the $q_{-}=1$, $z_{1}=2e^{2i\pi/3}$;  $\textbf{(d)}$: denotes the density graph.

As can be seen from the Figure 4, the propagation of the solution is neither parallel to the time axis nor parallel to the space axis, which is called the non-stationary soliton solution. By choosing special parameters, we can also obtain the space-periodic breather solution.

\noindent \textbf {Case 3.}
For $N=1$, $q_{-}=1$ and $z=e^{i\pi/4}$ means the  asymptotic phase difference is $\pi$, then the propagation behavior of the solution is shown in Figure 5.

{\rotatebox{0}{\includegraphics[width=2.85cm,height=2.8cm,angle=0]{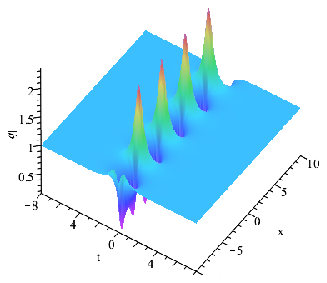}}}
{\rotatebox{0}{\includegraphics[width=2.35cm,height=2.15cm,angle=0]{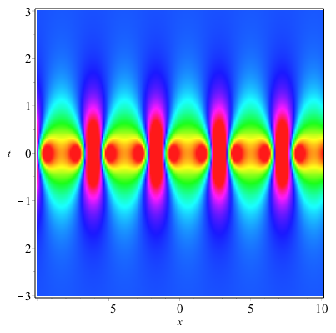}}}
{\rotatebox{0}{\includegraphics[width=2.85cm,height=2.8cm,angle=0]{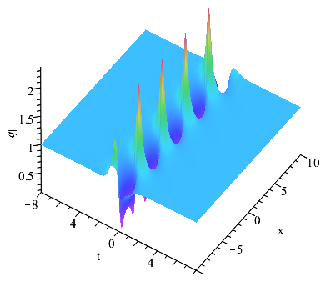}}}
{\rotatebox{0}{\includegraphics[width=2.35cm,height=2.15cm,angle=0]{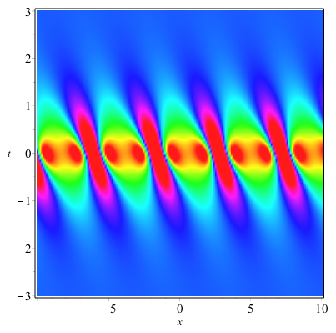}}}

 $\qquad\qquad(\textbf{a})\quad \ \qquad\qquad\qquad(\textbf{b})
\ \qquad\quad\quad\quad\qquad(\textbf{c})\qquad\qquad\qquad\qquad(\textbf{d})
\ \qquad$\\
\noindent { \small \textbf{Figure 5.} (Color online)  The breather wave solutions  with the fixed parameters $N=1$,  $C_{+}[z_{1}]=1$. Fig. $\textbf{(a)}$: the space-breather solution with the $q_{-}=1$, $z_{1}=e^{i\pi/4}$ and $\delta_{2}=0$; $\textbf{(b)}$: presents the density of the space-breather solution; Fig. $\textbf{(c)}$: the breather solution with the $q_{-}=1$, $z_{1}=e^{i\pi/4}$ and $\delta_{2}=1$; $\textbf{(d)}$: denotes the density of the breather solution.

Fig. $5$ shows that the solution of the equation changes periodically along the spatial direction ($x$-axis)  when there is no parameter influence ($\delta_{2}=0$), that is, the so-called space-periodic breather solution. When the modulus of the parameter ($\delta_{2}$) increases, the solution is no longer parallel to the $x$-axis, and the sign of the parameter only changes the propagation direction of the solution.

\noindent \textbf {Case 4.}
When $N=2$, we will show the weak interaction of three different state solutions with some special parameters.

{\rotatebox{0}{\includegraphics[width=3.6cm,height=3.0cm,angle=0]{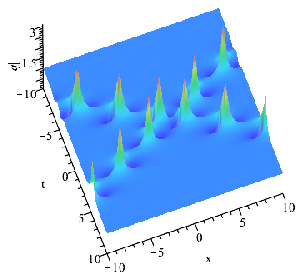}}}
{\rotatebox{0}{\includegraphics[width=3.3cm,height=2.55cm,angle=0]{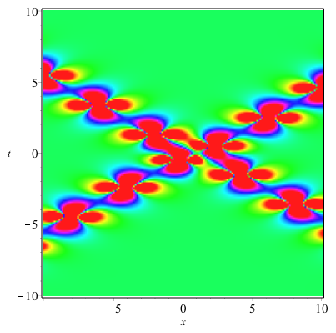}}}
{\rotatebox{0}{\includegraphics[width=3.6cm,height=2.75cm,angle=0]{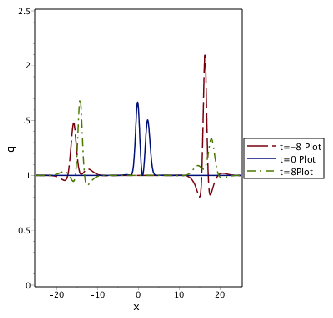}}}

$\qquad\qquad\quad(\textbf{a1})\quad \ \quad\qquad\qquad\qquad\qquad(\textbf{a2})
\qquad\qquad\qquad\quad\quad(\textbf{a3})$\\

{\rotatebox{0}{\includegraphics[width=3.6cm,height=3.0cm,angle=0]{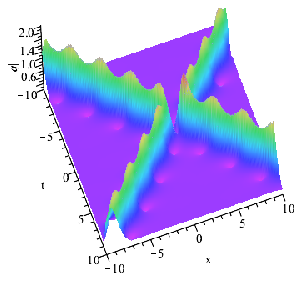}}}
{\rotatebox{0}{\includegraphics[width=3.3cm,height=2.55cm,angle=0]{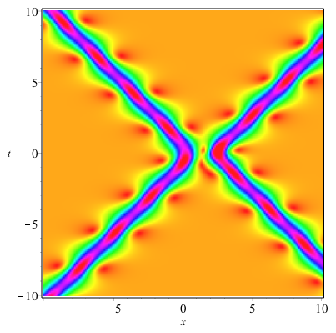}}}
{\rotatebox{0}{\includegraphics[width=3.6cm,height=2.75cm,angle=0]{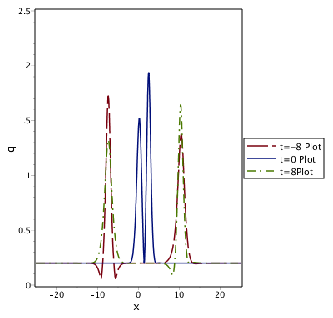}}}

$\qquad\qquad\quad(\textbf{b1})\quad \ \quad\qquad\qquad\qquad\qquad(\textbf{b2})
\qquad\qquad\qquad\quad\quad(\textbf{b3})$\\

{\rotatebox{0}{\includegraphics[width=3.6cm,height=3.0cm,angle=0]{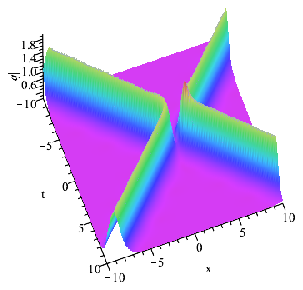}}}
{\rotatebox{0}{\includegraphics[width=3.3cm,height=2.55cm,angle=0]{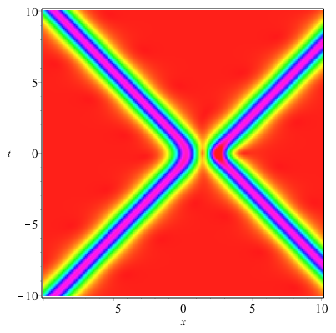}}}
{\rotatebox{0}{\includegraphics[width=3.6cm,height=2.75cm,angle=0]{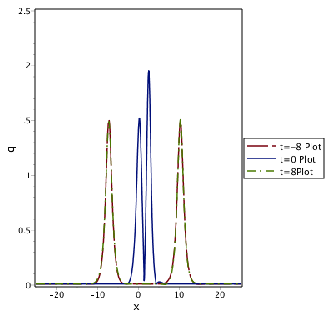}}}

$\qquad\qquad\quad(\textbf{c1})\quad \ \quad\qquad\qquad\qquad\qquad(\textbf{c2})
\qquad\qquad\qquad\quad\quad(\textbf{c3})$\\

\noindent { \small \textbf{Figure 6.} (Color online) Plots of the breather solution with the parameters $C_{+}[z_{1}]=C_{+}[z_{2}]=1$, $Z_{1}=\frac{1}{2}+\frac{3}{2}i$, $Z_{2}=-\frac{1}{2}+\frac{3}{2}i$ and $\delta_{2}=0$.
$\textbf{(a1, a2, a3)}$ show the breather-breather solution with $q_{-}=1$;
$\textbf{(b1, b2, b3)}$ present the breather-breather solution with $q_{-}=0.2$;
$\textbf{(c1,c2,c3)}$ display the bright-bright solution with $q_{-}=0.01$.\\

Similar to the previous case $1$, the weak interaction between the breather solution and the breather solution (see  $Fig. 6(a1)-6(a3)$)  gradually changes into the interaction between bright solitons (see $Fig. 6(c1)-6(c3)$) with the decrease of the boundary conditions $q_{-}$, and the periodic behavior of the solution gradually moves up and occurs at the top.

\noindent \textbf {Case 5.}
On the basis of Case $4$, we study the influence of parameters $\delta_{2}$ on solution propagation behavior when $N=2$

{\rotatebox{0}{\includegraphics[width=2.75cm,height=2.5cm,angle=0]{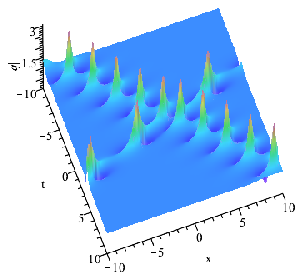}}}
{\rotatebox{0}{\includegraphics[width=2.75cm,height=2.5cm,angle=0]{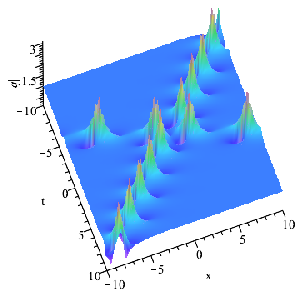}}}
{\rotatebox{0}{\includegraphics[width=2.75cm,height=2.5cm,angle=0]{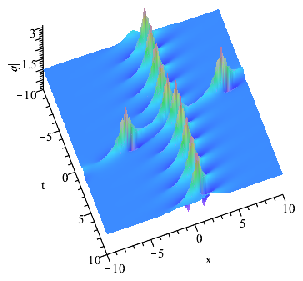}}}
{\rotatebox{0}{\includegraphics[width=2.75cm,height=2.5cm,angle=0]{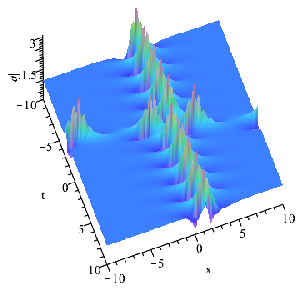}}}

 $\qquad\qquad(\textbf{a1})\qquad \ \qquad\quad\qquad(\textbf{b1})
\ \qquad\quad\quad\quad\qquad(\textbf{c1})\qquad\qquad\quad\qquad(\textbf{d1})
\ \qquad$\\

{\rotatebox{0}{\includegraphics[width=2.75cm,height=2.5cm,angle=0]{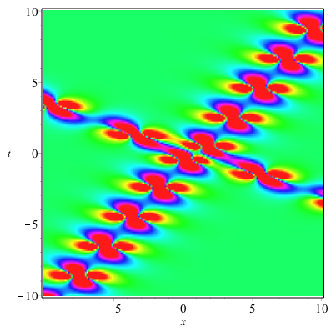}}}
{\rotatebox{0}{\includegraphics[width=2.75cm,height=2.5cm,angle=0]{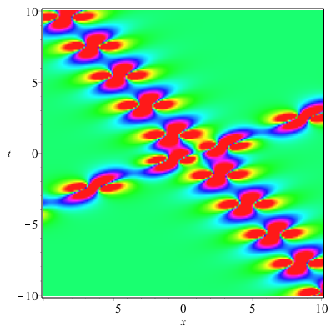}}}
{\rotatebox{0}{\includegraphics[width=2.75cm,height=2.5cm,angle=0]{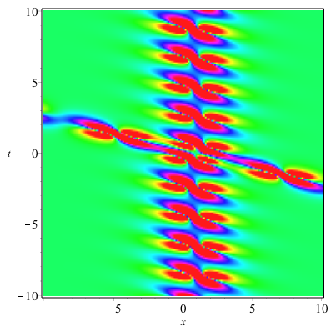}}}
{\rotatebox{0}{\includegraphics[width=2.75cm,height=2.5cm,angle=0]{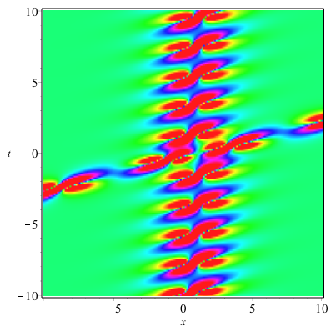}}}

 $\qquad\quad\quad(\textbf{a2})\qquad \ \qquad\quad\qquad(\textbf{b2})
\ \qquad\quad\quad\quad\qquad(\textbf{c2})\qquad\qquad\quad\qquad(\textbf{d2})
\ \qquad$\\

\noindent { \small \textbf{Figure 7.} (Color online)  Plots of the breather solution with the parameters $C_{+}[z_{1}]=C_{+}[z_{2}]=1$, $z_{1}=\frac{1}{2}+\frac{3}{2}i$, $z_{2}=-\frac{1}{2}+\frac{3}{2}i$. $\textbf{(a1, a2)}$: the breather solution with the $\delta_{2}=1$; $\textbf{(b1, b2)}$: the breather solution with the $\delta_{2}=-1$; $\textbf{(c1, c2)}$: the breather solution with the $\delta_{2}=2$; $\textbf{(d1, d2)}$: the breather solution with the $\delta_{2}=-2$;

Contrast Fig. $7(a1,a2)$  and Fig. $7(b1,b2)$ show that the positive and negative sign of parameters affect the direction of solution propagation without changing the shape and size of solution; compare Fig. $7(c1,c2)$  with Fig. $7(d1,d2)$  exhibit that when the parameters $\delta_{2}$ increase gradually, the shape of solution changes and the shape becomes irregular gradually.  From Fig. $6(a1,a2)$, with the parameters $\delta_{2}$ is of zero, the energy of the two breather solution is uniform and distributed when they interact. However from Fig. $7(a1,a2)$, we know that when the parameter $\delta_{2}\neq0$, the energy transfer of the two breather solution will occur during the interaction.
\section{NZBCs with the double poles}
\subsection{Residue conditions}
Based on the previous work, we further discuss the case of the double poles of the equation under non-zero boundary conditions. This process is similar to a single pole, but there are many differences, including the form of the solution, theta condition, etc. In what following, we consider that  the discrete spectral $Z^{d}$ is a double zero of the scattering coefficient, which means that $z_{0}$ is a double zero of $s_{11}(z)$, then $s_{11}(z_{0})=s'_{11}(z_{0})=0$ along with $s''_{11}(z_{0})\neq0$ (Note here and in what follows $'$ denotes the derivative with respect to $z$). In the latter calculation we need to use the following proposition about residue condition \cite{Pichler-2017}

\noindent \textbf {Proposition 8.}
\emph{If the function $f$ and $g$ are analytic in a  complex  region $\Omega\in\mathbb{C}$ with $g$ is of a double zero $z_{0}\in\Omega$ and $f(z_{0}\neq0)$. Thus the residue condition of $f/g$ can be derived by the Laurent expansion at $z=z_{0}$, namely}
\begin{align}\label{T1}
\mathop{Res}_{z=z_{0}}\left[\frac{f}{g}\right]=\frac{2f'(z_{0})}{g''(z_{0})}-
\frac{2f(z_{0})g'''(z_{0})}{3(g''(z_{0}))^{2}},\quad
\mathop{P_{-2}}_{z=z_{0}}\left[\frac{f}{g}\right]=\frac{2f(z_{0})}{g''(z_{0})}.
\end{align}

The discrete spectrum points discussed below are all double-pole cases. For all $z_{n}\in Z^{d}\cup D^{+}$, we have $s_{11}(z_{n})=s'_{11}(z_{n})=0$ along with $s''_{11}(z_{n})\neq0$. Recall that the first expression of \eqref{Q18}, there is exist a constant $b_{+}[z_{n}]$ (just use the same sign as before, the specific expressions are different) such that
\begin{align}\label{T2}
\begin{split}
\phi_{+,1}(x,t;z_{n})&=b_{+}[z_{n}]\phi_{-,2}(x,t;z_{n}),\\
\phi'_{+,1}(x,t;z_{n})&=d_{+}[z_{n}]\phi_{-,2}(x,t;z_{n})+
b_{+}[z_{n}]\phi'_{-,2}(x,t;z_{n}).
\end{split}
\end{align}
Similarly for all $z_{n}^{*}\in Z^{d}\cup D^{-}$, then $s_{11}(z_{n}^{*})=s'_{11}(z_{n}^{*})=0$ along with $s''_{11}(z_{n}^{*})\neq0$, one has
\begin{align}\label{T3}
\begin{split}
\phi_{+,2}(x,t;z_{n}^{*})&=b_{+}[z_{n}^{*}]\phi_{-,1}(x,t;z_{n}^{*}),\\
\phi'_{+,2}(x,t;z_{n}^{*})&=d_{+}[z_{n}^{*}]\phi_{-,1}(x,t;z_{n}^{*})+
b_{+}[z_{n}^{*}]\phi'_{-,1}(x,t;z_{n}^{*}).
\end{split}
\end{align}
Using the expression \eqref{T1} and \eqref{T2}, we first consider the residue condition of $u_{+,1}(x,t;z)/s_{11}(z)$:
\begin{align}\label{T4}
\begin{split}
\mathop{P_{-2}}_{z=z_{n}}\left[\frac{u_{+,1}(x,t;z)}{s_{11}(z)}\right]&=
\frac{2u_{+,1}(x,t;z_{n})}{s''_{11}(z_{n})}=
\frac{2b_{+}[z_{n}]}{s''_{11}(z_{n})}\phi_{-,2}(x,t;z_{n}),\\
\mathop{Res}_{z=z_{n}}\left[\frac{u_{+,1}(x,t;z)}{s_{11}(z)}\right]&=
\frac{2b_{+}[z_{n}]}{s''_{11}(z_{n})}\left[\phi'_{-,2}(x,t;z_{n})+
\left(\frac{d_{+}[z_{n}]}{b_{+}[z_{n}]}-\frac{s'''_{11}(z_{n})}{3s''_{11}(z_{n})}
\right)\phi_{-,2}(x,t;z_{n})\right].
\end{split}
\end{align}
For convenience, we denote $A_{+}[z_{n}]=\frac{2b_{+}[z_{n}]}{s''_{11}(z_{n})}$ and
$B_{+}[z_{n}]=\frac{d_{+}[z_{n}]}{b_{+}[z_{n}]}-
\frac{s'''_{11}(z_{n})}{3s''_{11}(z_{n})}$.
Similarly, we can derive the residual expression of $u_{+,2}(x,t;z)/s_{22}(z)$ for all $z_{n}^{*}\in Z^{d}\cup D^{-}$:
\begin{align}\label{T5}
\begin{split}
\mathop{P_{-2}}_{z=z_{n}^{*}}\left[\frac{u_{+,2}(x,t;z)}{s_{22}(z)}\right]&=
\frac{2u_{+,2}(x,t;z_{n}^{*})}{s''_{22}(z_{n}^{*})}=
\frac{2b_{-}[z_{n}^{*}]}{s''_{22}(z_{n}^{*})}\phi_{-,1}(x,t;z_{n}^{*}),\\
\mathop{Res}_{z=z_{n}^{*}}\left[\frac{u_{+,2}(x,t;z)}{s_{22}(z)}\right]&=
\frac{2b_{-}[z_{n}^{*}]}{s''_{22}(z_{n}^{*})}\left[\phi'_{-,1}(x,t;z_{n}^{*})+
\left(\frac{d_{-}[z_{n}^{*}]}{b_{-}[z_{n}^{*}]}-\frac{s'''_{22}(z_{n}^{*})}{3s''_{22}(z_{n}^{*})}
\right)\phi_{-,1}(x,t;z_{n}^{*})\right].
\end{split}
\end{align}
Also we take $A_{-}[z_{n}^{*}]=\frac{2b_{-}[z_{n}^{*}]}{s''_{22}(z_{n}^{*})}$ and
$B_{-}[z_{n}^{*}]=\frac{d_{-}[z_{n}^{*}]}{b_{-}[z_{n}^{*}]}-
\frac{s'''_{22}(z_{n}^{*})}{3s''_{22}(z_{n}^{*})}$. Further the expression of \eqref{T4} and \eqref{T5} can be written as
\begin{align}
\mathop{P_{-2}}_{z=z_{n}}\left[\frac{u_{+,1}(x,t;z)}{s_{11}(z)}\right]&=
A_{+}[z_{n}]\phi_{-,2}(x,t;z_{n}),\\
\mathop{Res}_{z=z_{n}}\left[\frac{u_{+,1}(x,t;z)}{s_{11}(z)}\right]&=
A_{+}[z_{n}]\left[\phi'_{-,2}(x,t;z_{n})+B_{+}[z_{n}]\phi_{-,2}(x,t;z_{n})\right],\\
\mathop{P_{-2}}_{z=z_{n}^{*}}\left[\frac{u_{+,2}(x,t;z)}{s_{22}(z)}\right]&=
A_{-}[z_{n}^{*}]\phi_{-,1}(x,t;z_{n}^{*}),\\
\mathop{Res}_{z=z_{n}^{*}}\left[\frac{u_{+,2}(x,t;z)}{s_{22}(z)}\right]&=
A_{-}[z_{n}^{*}]\left[\phi'_{-,1}(x,t;z_{n}^{*})+
B_{-}[z_{n}^{*}]\phi_{-,1}(x,t;z_{n}^{*})\right].
\end{align}
Using the relation \eqref{Sy-1} and \eqref{Sy-2}, one has the symmetries (Please refer to \cite{Pichler-2017,Yan-2019} for details)
\begin{align}
\left\{
\begin{aligned}
A_{+}[z_{n}]&=-A_{-}^{*}[z_{n}^{*}]=\frac{z_{n}^{4}q_{-}^{*}}
{q_{0}^{4}q_{-}}A_{-}[-q_{0}^{2}/z_{n}]=-\frac{z_{n}^{4}q_{-}^{*}}
{q_{0}^{4}q_{-}}A_{+}[-q_{0}^{2}/z_{n}^{*}],\quad\quad z_{n}\in Z_{d}\cap D^{+},\\
B_{+}[z_{n}]&=B_{-}^{*}[z_{n}^{*}]=\frac{q_{0}^{2}}
{z_{n}^{2}}B_{-}[-q_{0}^{2}/z_{n}]+\frac{2}{z_{n}}=\frac{q_{0}^{2}}
{z_{n}^{2}}B_{+}^{*}[-q_{0}^{2}/z_{n}^{*}]+\frac{2}{z_{n}},
\quad z_{n}\in Z_{d}\cap D^{-}.
\end{aligned}
\right.
\end{align}
Then from \eqref{TE}, \eqref{F3}, \eqref{T4} and \eqref{T5}, the following residue conditions can be derived
\begin{align*}
\mathop{P_{-2}}_{z=\xi_{n}}M_{1}^{+}(x,t;z)&=\mathop{P_{-2}}_{z=\xi_{n}}
\left[\frac{u_{+,1}(x,t;z)}{s_{11}(z)}\right]=A[\xi_{n}]e^
{-2i\theta(x,t;\xi_{n})}u_{-,2}(x,t;\xi_{n}),\\
\mathop{P_{-2}}_{z=\hat{\xi}_{n}}M_{2}^{-}(x,t;z)&=\mathop{P_{-2}}_{z=\hat{\xi}_{n}}
\left[\frac{u_{+,2}(x,t;z)}{s_{22}(z)}\right]=A[\hat{\xi}_{n}]e^
{2i\theta(x,t;\hat{\xi}_{n})}u_{-,1}(x,t;\hat{\xi}_{n}),\\
\mathop{Res}_{z=\xi_{n}}M_{1}^{+}(x,t;z)&=\mathop{Res}_{z=\xi_{n}}
\left[\frac{u_{+,1}(x,t;z)}{s_{11}(z)}\right]=A[\xi_{n}]e^
{-2i\theta(x,t;\xi_{n})}\left\{u'_{-,2}(x,t;\xi_{n})+\left[B[\xi_{n}]-
2i\theta'(x,t;\xi_{n})\right]u_{-,2}(x,t;\xi_{n})\right\},\\
\mathop{Res}_{z=\hat{\xi}_{n}}M_{2}^{-}(x,t;z)&=\mathop{Res}_{z=\hat{\xi}_{n}}
\left[\frac{u_{+,2}(x,t;z)}{s_{22}(z)}\right]=A[\hat{\xi}_{n}]e^
{-2i\theta(x,t;\hat{\xi}_{n})}\left\{u'_{-,1}(x,t;\xi_{n})+\left[B[\hat{\xi}_{n}]+
2i\theta'(x,t;\hat{\xi}_{n})\right]u_{-,1}(x,t;\hat{\xi}_{n})\right\}.
\end{align*}
\subsection{RH problem with double poles}
The RH problem established by Proposition $6$ is still true for the case of double poles, that is
\begin{align}\label{T6}
M^{-}(x,t;z)=M^{+}(x,t;z)(\mathbb{I}-G(x,t;z)),
\end{align}
and asymptotic behavior, the jump matrix $G(x,t;z)$ are the same as before. Thus to solve the regularize RH problem \eqref{T6}, we also subtract  the asymptotic value \eqref{jianjin} and the singular contributions from the poles $Z^{d}$, then
\begin{align}\label{T7}
\begin{split}
 M^{-}-&\mathbb{I}-\frac{i}{z}\sigma_{3}Q_{-}-\sum_{n=1}^{2N}\left\{
\frac{\mathop{Res}\limits_{z=\xi_{n}}M^{+}}{z-\xi_{n}}
+\frac{\mathop{P_{-2}}\limits_{z=\xi_{n}}M^{+}}{(z-\xi_{n})^{2}}
+\frac{\mathop{Res}\limits_{z=\hat{\xi}_{n}}M^{-}}{z-\hat{\xi}_{n}}
+\frac{\mathop{P_{-2}}\limits_{z=\hat{\xi}_{n}}M^{+}}{(z-\hat{\xi}_{n})^{2}}\right\}\\&=
M^{+}-\mathbb{I}-\frac{i}{z}\sigma_{3}Q_{-}-\sum_{n=1}^{2N}\left\{
\frac{\mathop{Res}\limits_{z=\xi_{n}}M^{+}}{z-\xi_{n}}
+\frac{\mathop{P_{-2}}\limits_{z=\xi_{n}}M^{+}}{(z-\xi_{n})^{2}}
+\frac{\mathop{Res}\limits_{z=\hat{\xi}_{n}}M^{-}}{z-\hat{\xi}_{n}}
+\frac{\mathop{P_{-2}}\limits_{z=\hat{\xi}_{n}}M^{+}}{(z-\hat{\xi}_{n})^{2}}\right\}
-M^{+}G.
\end{split}
\end{align}
Based on the Cauchy projectors and Plemelj's formulae, the RH problem \eqref{T7} can be determined by
\begin{align}\label{jie-d}
\begin{split}
M(x,t;z)=\mathbb{I}+\frac{i}{z}\sigma_{3}Q_{-}&+\sum_{n=1}^{2N}\left\{
\frac{\mathop{Res}\limits_{z=\xi_{n}}M^{+}}{z-\xi_{n}}
+\frac{\mathop{P_{-2}}\limits_{z=\xi_{n}}M^{+}}{(z-\xi_{n})^{2}}
+\frac{\mathop{Res}\limits_{z=\hat{\xi}_{n}}M^{-}}{z-\hat{\xi}_{n}}
+\frac{\mathop{P_{-2}}\limits_{z=\hat{\xi}_{n}}M^{+}}{(z-\hat{\xi}_{n})^{2}}\right\}\\
&+\frac{1}{2i\pi}\int_{\Sigma}\frac{M(x,t;\zeta)G(x,t;\zeta)}{\zeta-z}\,d\zeta,\quad
z\in\mathbb{C}\setminus\Sigma,
\end{split}
\end{align}
\emph{where $\int_{\Sigma}$ means the counter shown in Fig. 1 (right).}
\subsection{Reconstruction formula for the potential with double poles}
\noindent \textbf {Proposition 9.}
\emph{The potential with double poles of the modified Landau-Lifshitz equation  is defined by}
\begin{align}\label{JIE-Z}
q(x,t)=q_{-}-i\sum_{n=1}^{2N}A_{-}[\hat{\xi}_{n}]e^{2i\theta(\hat{\xi}_{n})}(
u'_{-,1,1}(x,t;\hat{\xi}_{n})+\hat{D}_{n}u_{-,1,1}(x,t;\hat{\xi}_{n}))
+\frac{1}{2\pi}\int_{\Sigma}(M^{+}G)_{12}(x,t;\xi)\,d\xi.
\end{align}
\begin{proof}
Considering the second column value of \eqref{JIE-Z} at the discrete spectral point $z=\xi_{s}$ $s=1,2,\cdots,2N$, we get
\begin{align}\label{T9}
u_{-,2}(z)=\left(\begin{array}{cc}
                       \frac{iq_{-}}{z} \\
                        1
                     \end{array}\right)
+\sum_{k=1}^{2N}\hat{C}_{n}(z)
\left[u'_{-,1}(\hat{\xi}_{n})+\left(\hat{D}_{n}+\frac{1}{z-\hat{\xi}_{n}}\right)
u_{-,1}(\hat{\xi}_{n})\right]
+\frac{1}{2i\pi}\int_{\Sigma}\frac{(M^{+}G)_{2}(\xi)}{\xi-z}\,d\xi.
\end{align}
Substituting  the symmetry \eqref{Sy-2} into the left-hand of \eqref{T9}, one can get the expression only about $u_{-,1}$ with $z=\xi_{s}$ ($s=1,2,\cdots,2N$) as follow
\begin{align}\label{T10}
\sum_{k=1}^{2N}\hat{C}_{n}(\xi_{s})u'_{-,1}(\hat{\xi}_{n})+
\left[\hat{C}_{n}(\xi_{k})\left(\hat{D}_{n}+\frac{1}{\xi_{s}-\hat{\xi}_{n}}\right)
-\frac{iq_{-}}{\xi_{s}}\delta_{sn}\right]u_{-,1}(\hat{\xi}_{n})=-\left(\begin{array}{cc}
                       \frac{iq_{-}}{z} \\
                        1
                     \end{array}\right)
-\frac{1}{2i\pi}\int_{\Sigma}\frac{(M^{+}G)_{2}(\xi)}{\xi-\xi_{k}}\,d\xi.
\end{align}
Here $\delta_{sn}$ is the kronecker delta and
\begin{align}
C_{n}(z)=\frac{A_{+}[\xi_{n}]}{z-\xi_{n}}e^{-2i\theta(\xi_{n})},\quad
D_{n}=B_{+}[\xi_{n}]-2i\theta'(\xi_{n});\\
\hat{C}_{n}(z)=\frac{A_{-}[\hat{\xi}_{n}]}{z-\hat{\xi}_{n}}
e^{2i\theta(\hat{\xi}_{n})},\quad
\hat{D}_{n}=B_{-}[\hat{\xi}_{n}]+2i\theta'(\hat{\xi}_{n}).
 \end{align}
 Further determination of the value for the derivative of $u_{-,2}$ with respect to $z$ at the discrete spectral point $z=\xi_{s}$, we have
\begin{align}\label{T11}
u'_{-,2}(z)=\left(\begin{array}{cc}
                       -\frac{iq_{-}}{z^{2}} \\
                        0
                     \end{array}\right)
-\sum_{k=1}^{2N}\frac{\hat{C}_{n}(z)}{z-\hat{\xi}_{n}}
\left[u'_{-,1}(\hat{\xi}_{n})+\left(\hat{D}_{n}+\frac{2}{z-\hat{\xi}_{n}}\right)
u_{-,1}(\hat{\xi}_{n})\right]
+\frac{1}{2i\pi}\int_{\Sigma}\frac{(M^{+}G)_{2}(\xi)}{(\xi-z)^{2}}\,d\xi.
\end{align}
What's more, taking the derivative with respect to $z$ about \eqref{Sy-2}, the relationship can be obtianed
\begin{align}\label{T12}
u'_{-,2}(z)=-\frac{iq_{-}}{z^{2}}u_{-,1}(-q_{0}^{2}/z)+
\frac{iq_{0}^{2}q_{-}}{z^{3}}u'_{-,1}(-q_{0}^{2}/z).
\end{align}
Then plugging the \eqref{T12} into \eqref{T11}, yielding
\begin{align}\label{T13}
\begin{split}
\sum_{k=1}^{2N}\left(\frac{\hat{C}_{\xi_{s}}}{\xi_{s}-\hat{\xi}_{n}}
+\frac{iq_{0}^{2}q_{-}}{\xi_{s}^{3}}\right)u'_{-,1}(\hat{\xi}_{n})+
\left[\frac{\hat{C}_{n}(\xi_{s})}{\xi_{s}-\hat{\xi}_{n}}
\left(\hat{D}_{n}+\frac{2}{\xi_{s}-\hat{\xi}_{n}}\right)
-\frac{iq_{-}}{\xi_{s}^{2}}\delta_{sn}\right]u_{-,1}(\hat{\xi}_{n})\\
=\left(\begin{array}{cc}
                       -\frac{iq_{-}}{\xi_{s}^{2}} \\
                        0
                     \end{array}\right)
+\frac{1}{2i\pi}\int_{\Sigma}\frac{(M^{+}G)_{2}(\xi)}{(\xi-\xi_{k})^{2}}\,d\xi.
\end{split}
\end{align}
Similar to the process of proving theorem $5$,  because of the asymptotic behavior of $M(x,t;z)$ along with the Taylor expansion, we know that
\begin{align}\label{T15}
\begin{split}
M^{(1)}(x,t;z)&=i\sigma_{3}Q_{-}-\frac{1}{2\pi i}\int_{\Sigma}M^{+}(x,t;\zeta)G(x,t;\zeta)\,d\zeta\\&+
\sum_{n=1}^{2N}[A_{+}[\xi_{n}]e^{-2i\theta(\xi_{n})}(u'_{-,2}(\xi_{n})
+D_{n}u_{-,2}(\hat{\xi}_{n})),
A_{-}[\hat{\xi}_{n}]e^{2i\theta(\hat{\xi}_{n})}(u'_{-,1}(\hat{\xi}_{n})
+\hat{D}_{n}u_{-,1}(\hat{\xi}_{n}))].
\end{split}
\end{align}
Further, by comparing the coefficients $z^{0}$, the expression of the final solution $q(x,t)$ can be written in the form of propositions 9.
\end{proof}
\subsection{Trace formula and theta condition}
The theta condition and the trace formula of double-poles are different from those of simple point. Next, we will deduce the trace formula of double-poles point. That is to say, scattering coefficient $s_{11}(z)$ and $s_{22}(z)$ are expressed by discrete eigenvalue and reflection coefficient. When the discrete spectral $z_{n}$ and $-q_{0}^{2}/z_{n}$ are the double zeros of $s_{11}(z)$, then the function
\begin{align}\label{T16}
\Xi^{+}(z)=s_{11}(z)\prod_{n=1}^{2N}\frac{(z-z_{n}^{*})^{2}(z+q_{0}^{2}/z_{n})^{2}}
{(z-z_{n})^{2}(z+q_{0}^{2}/z_{n}^{*})^{2}}
\end{align}
is analytic in $D^{+}$ as well as there is no zeros in the region. Similarly
\begin{align}\label{T17}
\Xi^{-}(z)=s_{22}(z)\prod_{n=1}^{2N}\frac{(z-z_{n})^{2}(z+q_{0}^{2}/z_{n}^{*})^{2}}
{(z-z_{n}^{*})^{2}(z+q_{0}^{2}/z_{n})^{2}}
\end{align}
is analytic in $D^{-}$ as well as there is no zeros in the region. obviously $\Xi^{+}(z)\Xi^{-}(z)=s_{11}(z)s_{22}(z)$ on $\Sigma$. And \eqref{T14} implies
\begin{align}\label{T18}
\Xi^{+}(z)\Xi^{-}(z)=\frac{1}{1-\rho(z)\tilde{\rho}(z)},\quad z\in\Sigma.
\end{align}
Similar to the previous process, taking logarithm and  using Cauchy operator, we obtain
\begin{align}
s_{11}(z)&=exp\left(-\frac{1}{2\pi i }\frac{\log[1-\rho(\zeta)\tilde{\rho}(\zeta)]}{\zeta-z}
\,d\zeta\right)\prod_{n=1}^{2N}\frac{(z-z_{n})^{2}(z+q_{0}^{2}/z_{n}^{*})^{2}}
{(z-z_{n}^{*})^{2}(z+q_{0}^{2}/z_{n})^{2}},\\
s_{22}(z)&=exp\left(-\frac{1}{2\pi i }\frac{\log[1-\rho(\zeta)\tilde{\rho}(\zeta)]}{\zeta-z}
\,d\zeta\right)\prod_{n=1}^{2N}\frac{(z-z_{n}^{*})^{2}(z+q_{0}^{2}/z_{n})^{2}}
{(z-z_{n})^{2}(z+q_{0}^{2}/z_{n}^{*})^{2}}.
\end{align}
Finally, the so-called theta condition, i.e., the asymptotic phase difference that determines the boundary value, can be expressed by scattering coefficients.
\begin{align}
\arg\frac{q_{-}}{q_{+}}=argq_{-}-argq_{+}=\frac{1}{2\pi}\int_{\Sigma}
\frac{\log[1-\rho(\zeta)\tilde{\rho}(\zeta)]}{\zeta}\,d\zeta+8\sum_{n=1}^{N}\arg z_{n}.
\end{align}
\subsection{Soliton solutions with double poles}
\noindent \textbf {Proposition 10.}
The solution for the reflection-ness of the modified Landau-Lifshitz equation with double poles can be written as
\begin{align}\label{T19}
q(x,t)=q_{-}+i\frac{\det\left(
                \begin{array}{cc}
                  M & \nu \\
                  \mu & 0 \\
                \end{array}
              \right)}{\det M},
              \end{align}
here $M=\left(\begin{array}{cc}
                  M^{(11)} & M^{(12)} \\
                  M^{(21)} & M^{(22)} \\
                \end{array}
              \right)$  and $M^{(ij)}=(m_{kn}^{(ij)})_{(2N)\times(2N)}$ with
 $m_{kn}^{(11)}=\hat{C}_{n}(\xi_{k})\left(\hat{D}_{n}+\frac{1}{\xi_{k}-
 \hat{\xi}_{n}}\right)-\frac{iq_{-}}{\xi_{k}}\delta_{k,n}$, $m_{kn}^{(12)}=\hat{C}_{n}(\xi_{k})$,
 $m_{kn}^{(21)}=\frac{\hat{C}_{n}(\xi_{k})}{\xi_{k}-\hat{\xi}_{n}}\left(\hat{D}_{n}
 +\frac{2}{\xi_{k}-\hat{\xi}_{n}}\right)-\frac{iq_{-}}
 {\xi_{k}^{2}}\delta_{k,n}$,
 $m_{kn}^{(22)}=\frac{\hat{C}_{n}(\xi_{k})}{\xi_{k}-\hat{\xi}_{n}}+
 \frac{iq_{-}q_{0}^{2}}{\xi_{k}^{3}}\delta_{k,n}$
$\mu_{n}^{(1)}=A_{-}[\hat{\xi_{n}}]e^{2i\theta(\hat{\xi_{n}})}\hat{D}_{n}$,
$\mu_{n}^{(2)}=A_{-}[\hat{\xi_{n}}]e^{2i\theta(\hat{\xi_{n}})}$,
$\nu_{n}^{(1)}=-\frac{iq_{-}}{\xi_{n}}$, $\nu_{n}^{(2)}=-\frac{iq_{-}}{\xi_{n}^{2}}$.

In order to further explain the properties of the solution \eqref{T19}, we select appropriate parameters to draw the propagation behavior of the solution using Maple, and analyze the influence of parameters on its propagation.

{\rotatebox{0}{\includegraphics[width=2.75cm,height=2.5cm,angle=0]{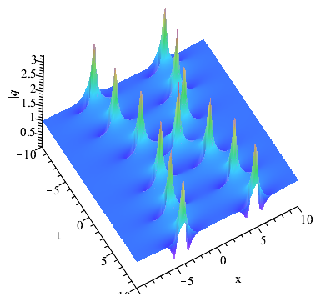}}}
{\rotatebox{0}{\includegraphics[width=2.75cm,height=2.5cm,angle=0]{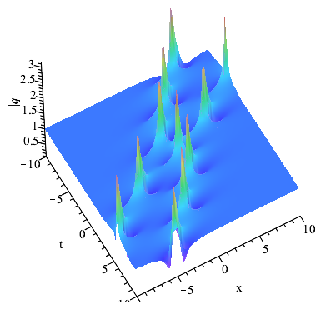}}}
{\rotatebox{0}{\includegraphics[width=2.75cm,height=2.5cm,angle=0]{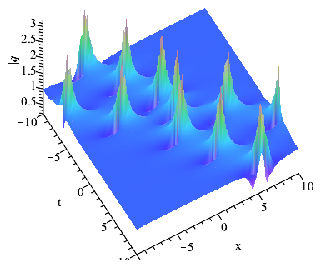}}}
{\rotatebox{0}{\includegraphics[width=2.75cm,height=2.5cm,angle=0]{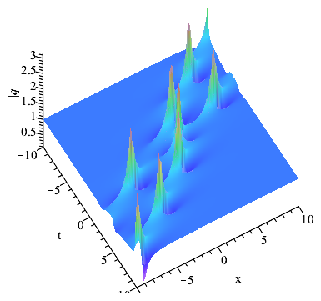}}}

 $\qquad\qquad(\textbf{a})\qquad \ \qquad\qquad\qquad(\textbf{b})
\ \qquad\quad\quad\qquad\qquad(\textbf{c})\qquad\qquad\qquad\qquad(\textbf{d})
\ \qquad$\\
\noindent { \small \textbf{Figure 8.} (Color online)  The breather wave solutions  with the fixed parameters $N=1$, $z_{1}=3i/2$, $A_{+}[z_{1}]=B_{+}[z_{1}]1$ and $q_{-}=1$. $\textbf{(a)}$: the breather solution with the $\delta_{2}=0$; $\textbf{(b)}$: the breather solution with the $\delta_{2}=1$; $\textbf{(c)}$: the breather solution with the $\delta_{2}=-1$; $\textbf{(d)}$: the breather solution with the $\delta_{2}=1.5$.

It can be seen from the Fig. $8$ that when the parameters $\delta_{2}=0$, the solution behaves as the interaction of two breather solutions, and when the parameters $\delta_{2}$ increase gradually, the propagation behavior of the solution becomes irregular; when the parameters $\delta_{2}$ is less than zero, the propagation direction of the solution changes, but does not change the shape and size of the solution.

\section{Conclusions and discussions}
In this work, the Cauchy problem of  the mLL equation
 with the non-zero boundary value is studied in detail based on generalized RH method. Due to the boundary value is not zero, some technical difficulties will arise, which leads to the difficulty of constructing RH problem. This problem is solved by means of Riemann surface. By next analyzing the asymptotic Lax pairs the Jost function, scattering matrix and their analytic symmetry are obtained, which are  need to be used to construct the RH problem in the inverse scattering process. In addition, the asymptotic analysis and the residue conditions and theta conditions at discrete spectral points are given. In the inverse scattering process, the generalized RH problem is given based on some data obtained from direct scattering.  The solutions of the mLL equation, such as time-periodic solution, space-periodic solution, non-stationary solution, bright soliton solution etc, are obtained by solving the RH problem. Finally we discuss the double poles case, at which time the solution of the mLL equation, the trace formula, the theta condition etc have changed.
  Appropriate parameters are selected to discuss the solutions of different parameters and the influence of propagation behavior includes whether it affects the shape, size and direction of the solution.
\section*{Acknowledgements}
the Postgraduate Research and Practice of Educational Reform for Graduate students in CUMT under Grant No. 2019YJSJG046, the Natural Science Foundation of Jiangsu Province under Grant No. BK20181351, the Six Talent Peaks Project in Jiangsu Province under Grant No. JY-059, the Qinglan Project of Jiangsu Province of China, the National Natural Science Foundation of China under Grant Nos. 11975306 and 61877053, the Fundamental Research Fund for the Central Universities under the Grant Nos. 2019ZDPY07 and 2019QNA35, and the General Financial Grant from the China Postdoctoral Science Foundation under Grant Nos. 2015M570498 and 2017T100413.

\end{document}